\documentclass[aps,showpacs,twocolumn,superscriptaddress]{revtex4}
\usepackage{amsmath,amssymb}
\usepackage{amsmath}
\usepackage{graphicx}

\begin{document}

\title{Coherent charge transport through molecular wires: \textquotedblleft
Exciton blocking\textquotedblright and current from electronic excitations
in the wire}
\date{\today }

\author{GuangQi Li}
\affiliation{Raymond and Beverly Sackler Faculty of Exact Sciences, School of Chemistry,
Tel-Aviv University, Tel-Aviv 69978, Israel}
\author{Boris D. Fainberg}
 \email{fainberg@hit.ac.il}
\affiliation{Raymond and Beverly Sackler Faculty of Exact Sciences, School of Chemistry,
Tel-Aviv University, Tel-Aviv 69978, Israel}
\affiliation{Faculty of Sciences, Holon Institute of Technology, 52 Golomb St., Holon
 58102, Israel}
 \author{Abraham Nitzan}
 \affiliation{Raymond and Beverly Sackler Faculty of Exact Sciences, School of Chemistry,
Tel-Aviv University, Tel-Aviv 69978, Israel}
\author{Sigmund Kohler}
\affiliation{Instituto de Ciencia de Materiales de Madrid (CSIC), Cantoblanco, 28049
 Madrid, Spain}
 \author{Peter H\"anggi}
 \affiliation{Institute for Physics, University of Augsburg,
Augsburg, D-86135, Germany}

\begin{abstract}
We consider exciton effects on current in molecular nanojunctions, using a
model comprising a two two-level sites bridge connecting free electron
reservoirs. Expanding the density operator in the many-electron eigenstates
of the uncoupled sites, we obtain a $16\times 16$ density matrix in the
bridge subspace whose dynamics is governed by Liuoville equation that takes
into account interactions on the bridge as well as electron injection and
damping to and from the leads. Our consideration can be considerably
simplified by using the pseudospin description based on the symmetry
properties of Lie group SU(2). We study the influence of the bias voltage,
the Coulomb repulsion and the energy-transfer interactions on the
steady-state current and in particular focus on the effect of the excitonic
interaction between bridge sites. Our calculations show that in case of
non-interacting electrons this interaction leads to reduction in the current
at high voltage for a homodimer bridge. In other words, we predict the
effect of \textquotedblleft exciton\textquotedblright blocking. The effect
of \textquotedblleft exciton\textquotedblright blocking is modified for a
heterodimer bridge, and disappears for strong Coulomb repulsion at sites. In
the latter case the exciton type interactions can open new channels for
electronic conduction. In particular, in the case of strong Coulomb
repulsion, conduction exists even when the electronic connectivity does not
exist.
\end{abstract}

\maketitle

\draft

\section{Introduction}

\label{sec:introduction}

Electron transport through molecular wires has been under intense
theoretical (see e.g. \cite{nitz03a,Nitzan08Science}) and experimental (see
e.g. \cite{Chen09,Heath09}) study in the last few years. Theoretical studies
usually fall into two categories. The first focuses on the \textit{ab-initio%
} computations of the orbitals relevant for the motion of excess charges
through the molecular wire \cite{Lang00,Schon02,Evers04,Ratner02CP,Ghosh02CP}%
, while the other \cite{Kohler05,Kaiser06} employs generic models to gain
qualitative understanding of the transport process. At the simplest level
\cite{Kohler05,Kaiser06} the wire Hamiltonian is described by a
tight-binding chain composed of $N$ sites with nearest-neighbor coupling
(Huckel model) that represents the electron transfer (tunneling)
interactions between adjacent sites. This model has been generalized to
include Coulomb interactions between electrons on the same site \cite{Meir92}
(Hubbard model) and/or electron-phonon interactions \cite{Nitzan07JPhys}. In
the present paper we investigate another extension of this model, in which
we take into account energy transfer interactions between adjacent molecular
sites.

Energy-transfer interactions - excitation (deexcitation) of a site
accompanied by deexcitation (excitation) of another are well-known in the
exciton theory \cite{Dav71,Agranovich68,Craig68}. In particular, Frenkel
excitons - neutral excited states in which an electron and a hole are placed
on the same site are readily transferred between sites, and such intersite
interactions can accompany the charge transfer processes as was shown for
charge-transfer excitons \cite{Agranovich77} in (quasi-) one-dimensional
structures \cite{Agranovich00,Reineker09}, including polysilanes \cite%
{Thorne90,Trommsdorff92,Shimizu00}. The latter show a weak coupling between
the Frenkel exciton with the admixture of charge transfer states and nuclear
motions \cite{Trommsdorff92,Shimizu00}.

In molecular bridges energy-transfer interactions can also sometimes have
important effects on charge transfer dynamics. Charge and energy transfer in
a linear 2,2':6',2\textquotedblright -terpyridine-based trinuclear
Ru-(II)-Os(II) nanometer-sized array \cite{BennistonJACS05}, and
one-dimensional energy/electron transfer of amylose-encapsulated chain
chromophores \cite{KimJACS06} are examples. In addition, it seems likely
that energy transfer takes place in chemically responsive molecular
transistors based on a dimer of terpyridyl molecules combined with ion Co$%
^{2+}$ \cite{Tang06}.

It should be noted that electron transfer is a tunneling process that
depends exponentially on the site-site distance, while energy transfer is
associated with dipolar coupling that scales like the inverse cube of this
distance, and can therefore dominate at larger distances. The importance of
the latter stems also from geometric issues, which are related to the
dipole-dipole interaction between different sites occurring in the vicinity
of metal particles in molecular nanojunctions. Really, Gersten and Nitzan
\cite{Gersten_NitzanCPL84,Gersten_NitzanJCP85} predicted accelerated
dipole-dipole energy transfer near a solid particle (see also \cite%
{Polman07,Polman09}), and in the last time a number of works devoted to the
exciton-plasmon interactions have been published \cite%
{Wiederrecht04,Wurtz07,Cade09,Bondarev09} that are related to physical
effects due to the local field enhancement \cite%
{Schatz03JPCB,Stockman03,Markel05,Wang_Shen06,Brixner06,Sukharev_Seideman07}.

How will such dipolar interactions affect the conduction properties of
molecular junctions? This question was addressed by Galperin, Nitzan and
Ratner by the example of a junction composed of one-site-wire and two metal
leads \cite{Nitzan06PRL}, where they predicted the existence of non-Landauer
current induced by the electron-hole excitations in the leads. To the best
of our knowledge, there were no analog treatment of simultaneous electron
and energy transfer (excitons) in multisite bridges. Here we address this
problem by using the Liouville-von Neumann equation (LNE) for the total
density operator to derive an expression for the conduction of a molecular
wire model that contains both electron and energy-transfer interactions.
While not a central issue of the present work, we note that energy transfer
is closely related to heat transfer through the molecular nanojunction - an
issue of important consequences for junction stability and integrity.

Treated separately, the simplest models of exciton and electron transport
may be represented by tight-binding transport models, albeit in different
representations. Indeed, in the wire Hamiltonian (see Eq.(\ref{eq:H_wire})
below), both the electron- and energy-transfer terms are binary in terms of
the annihilation and creation operators for electrons and excitons,
respectively. Their simultaneous treatment, however, constitutes a rather
complex non-linear problem. In this work we combine a tight-binding model
for electron transport \cite{Kohler05,Kaiser06} with that of one-dimensional
Frenkel excitons \cite{Dav71,Agranovich68,Craig68} to investigate the effect
of energy transfer interaction on electron transport in one-dimensional
nanowires. The outline of the paper is as follows. In Sec.\ref{sec:model} we
introduce our model and in Sec.\ref{sec:master_equation} we derive a master
equation in the eigenbasis of many-electron wire Hamiltonian. Sec.\ref%
{sec:Analytical} is devoted to the analytical solution of the problem where
we consider both non-interacting electrons at a site and strong Coulomb
repulsion at sites. In Sec.\ref{subsec:exciton_induced_current} we show that
the exciton type interactions can open new channels for electronic
conduction. In Sec.\ref{sec:Numerical simulations} we carry out numerical
simulations, compare them with the analytical theory and show the existence
of the \textquotedblleft exciton\ blocking\textquotedblright\ effect. We
summarize our results in Sec.\ref{sec:Conclusion}. In Appendix A we
calculate the eigenbasis of many-electron wire Hamiltonian for
non-interacting electrons at a site, using the Jordan-Wigner transformation
\cite{Fra98}. In Appendix B we present auxiliary calculations.

\section{Model}

\label{sec:model}

\begin{figure}[tbp]
\begin{center}
\includegraphics[width=7.cm,clip,angle=0]
{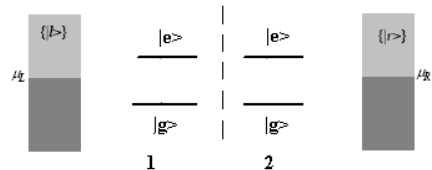}
\end{center}
\caption{A model for energy-transfer induced effects in molecular
conduction. The right ($R=|\{r\}\rangle $ ) and left ($L=|\{l\}\rangle $)
manifolds represent two metal leads characterized by electrochemical
potentials $\protect\mu _{R}$ and $\protect\mu _{L}$ respectively, each
coupled to its nearest molecular site. The molecular bridge is a dimer,
where each site is represented by its ground, $|1g\rangle $ and $|2g\rangle $%
, and excited, $|1e\rangle $ and$|2e\rangle $, states.}
\label{fig:dimer_model}
\end{figure}

We consider a spinless model for a molecular wire that comprises two
interacting sites, each represented by its ground, $|g\rangle $, and
excited, $|e\rangle $, states positioned between two leads represented by
free electron reservoirs $L$ and $R$ (Fig.\ref{fig:dimer_model}). The
electron reservoirs (leads) are characterized by their electronic chemical
potentials $\mu _{L}$ and $\mu _{R}$, where the difference $\mu _{L}-\mu
_{R} $ $=eV_{bs}$ is the imposed voltage bias. The corresponding Hamiltonian
is

\begin{equation}
\hat{H}=\hat{H}_{wire}+\hat{H}_{leads}+\hat{H}_{contacts}  \label{eq:Htotal}
\end{equation}%
\begin{equation}
\hat{H}_{leads}=\sum_{k\in \{L,R\}}\varepsilon _{k}\hat{c}_{k}^{+}\hat{c}_{k}
\label{eq:H_leads}
\end{equation}%
\begin{align}
\hat{H}_{wire}& =\sum_{\substack{ m=1,2  \\ f=g,e}}\varepsilon _{mf}\hat{c}%
_{mf}^{+}\hat{c}_{mf}-\sum_{f=g,e}\Delta _{f}(\hat{c}_{2f}^{+}\hat{c}_{1f}+%
\hat{c}_{1f}^{+}\hat{c}_{2f})+  \notag \\
& +\hbar J(b_{1}^{+}b_{2}+b_{2}^{+}b_{1})+\sum_{m=1,2}U_{m}N_{m}(N_{m}-1)
\label{eq:H_wire}
\end{align}%
\begin{equation}
\hat{H}_{contacts}=\hat{V}+\hat{W}  \label{eq:H_contacts}
\end{equation}

\begin{equation}
\hat{V}=\sum_{mf}\hat{V}_{mf}=\sum_{mf,k\in K_{m}}V_{k}^{(mf)}\hat{c}_{k}^{+}%
\hat{c}_{mf}+H.c.\text{,}  \label{eq:V_M1}
\end{equation}%
\begin{equation}
\hat{W}=\sum_{m}\hat{W}_{m}=\sum_{m,k\neq k^{\prime }\in K_{m}}\hat{W}%
_{kk^{\prime }}^{(m)}b_{k^{\prime }k}b_{n}^{+}+H.c.\text{,}
\label{eq:V_N_exciton2}
\end{equation}%
where $\hat{c}_{mf}^{+}$ ($\hat{c}_{mf}$) ($m=1,2$, $f=g,e$) are creation
(annihilation) operators for electrons in the different site states of
energies $\varepsilon _{mf}$, while $\hat{c}_{k}^{+}$ ($\hat{c}_{k}$) ($k\in
L,R$) are creation (annihilation) operators for free electrons (energies $%
\varepsilon _{k}$) in the leads $L$ and $R$. $\hat{n}_{mf}=\hat{c}_{mf}^{+}%
\hat{c}_{mf}$ are the occupation operators for the different site states,
and site occupation operators are given by $N_{m}=\hat{n}_{mg}+\hat{n}_{me}$%
. The operators $b_{m}^{+}=\hat{c}_{me}^{+}\hat{c}_{mg}$ and $b_{m}=\hat{c}%
_{mg}^{+}\hat{c}_{me}$ are excitonic (creation and annihilation) operators
on the molecular sites $m=1,2$, while $b_{k^{\prime }k}=\hat{c}_{k}^{+}\hat{c%
}_{k^{\prime }}=$ $b_{kk^{\prime }}^{+}$ ($k,k^{\prime }\in L$ or $R$ )
corresponds to electron-hole pairs in the leads. In the wire Hamiltonian,
Eq. (\ref{eq:H_wire}), the $\Delta _{f}$ terms represent electron hoping
between site states of similar energies (i.e. between $|g\rangle $ and
between $|e\rangle $ states of adjacent molecular sites), the $J$ terms
represent exciton hopping (energy transfer) between molecular sites and the $%
U$ terms correspond to on-site Coulomb interactions. The molecular-leads
interactions $\hat{H}_{contacts}$ are taken to account for two physical
processes: $\hat{V}$ describes electron transfer between the molecular
bridge and the leads that gives rise to net current in the biased junction,
while $\hat{W}$ describes energy transfer between the bridge and
electron-hole excitations in the leads. In (\ref{eq:V_M1}) and (\ref%
{eq:V_N_exciton2}) $K_{m}$ is the lead closer the the molecular site $m$ ($%
K_{1}=L$, $K_{2}=R$) and $H.c.$ denotes Hermitian conjugate. In what follows
it will be useful also to define the population operators
\begin{equation}
\lambda _{f}=\hat{n}_{2f}+\hat{n}_{1f}  \label{eq:lambda_f}
\end{equation}%
in the manifolds of ground ($f=g$ ) and excited ( $f=e$) site levels.

We consider electronic transport through the molecular wire where the leads $%
K=L,R$ are taken to be each in its own equilibrium characterized by its
temperature $T$ (here taken equal for the two leads) and electronic
electrochemical potential $\mu _{K}$. Therefore, the lead electrons are
described by the equilibrium Fermi functions $f_{K}(\varepsilon _{k})=[\exp
((\varepsilon _{k}-\mu _{K})/k_{B}T)+1]^{-1}$. Consequently expectation
values for lead operators can be traced back to the expression $\langle \hat{%
c}_{k}^{+}\hat{c}_{k^{\prime }}\rangle =f_{K}(\varepsilon _{k})\delta
_{kk^{\prime }}$ where $\delta _{kk^{\prime }}$ is the Kronecker delta. The
excitonic operators are equal to $b_{m}^{+}=\hat{c}_{me}^{+}\hat{c}_{mg}$.
The effect of the corresponding interaction in the bridge ($=\hbar
Jb_{1}^{+}b_{2}+H.c.$) on the charge transport properties is the subject of
our discussion.

\section{Master equation}

\label{sec:master_equation}

Our analysis is based on the LNE, or the generalized master equation for the
reduced density matrix of the molecular subsystem, obtained using a standard
procedure \cite{Kohler05,Kaiser06,Schreiber06} based on taking $\hat{H}%
_{contacts}$ as a perturbation. Briefly, one starts with the LNE for the
total density operator and use the projectors of the type $P_{K}\rho
(t)=\rho _{K}Tr_{K}\rho (t)$ in order to derive an equation for the time
evolution of the reduced density matrix $\sigma =Tr_{R}Tr_{L}\rho $. The
calculation is facilitated by invoking the so called non-crossing
approximation that assumes that the effects of different reservoirs (here $%
L,R$) and different relaxation processes (here $\hat{V}$, $\hat{W}$) are
independent and additive. This leads to

\begin{align}
\frac{d\sigma (t)}{dt}& =-\frac{i}{\hbar }[\hat{H}_{wire},\sigma (t)]-
\notag \\
& -\frac{1}{\hbar ^{2}}Tr_{K}\int_{0}^{\infty }dx[\hat{V},[\hat{V}%
^{int}(-x),\rho (t)]]  \notag \\
& -\frac{1}{\hbar ^{2}}Tr_{K}\int_{0}^{\infty }dx[\hat{W},[\hat{W}%
^{int}(-x),\rho (t)]]  \label{eq:density_matrix}
\end{align}%
where for any operator $\hat{O}$, $\hat{O}^{int}$ is the corresponding
interaction representation

\begin{align}
\hat{O}^{int}(-x)&=
\exp [-\frac{i}{\hbar }(\hat{H}_{wire}+\hat{H}_{leads})x] \nonumber \\
&\hat{O}\exp [\frac{i}{\hbar }(\hat{H}_{wire}+\hat{H}_{leads})x]
\label{eq:interaction represent}
\end{align}%
and where $Tr_{K}=Tr_{L}Tr_{R}$.

Consider first terms with the electron transfer interactions $\hat{V}$.
Writing the coupling Hamiltonians $\hat{V}_{nf}$ (Eq.(\ref{eq:V_M1})) as%
\begin{equation}
\hat{V}_{nf}=\hat{c}_{nf}\Lambda _{nf}^{+}+\hat{c}_{nf}^{+}\Lambda _{nf}
\label{eq:V_Mnf}
\end{equation}%
where $\Lambda _{nf}=\sum_{k\in K_{n}}V_{k}^{(nf)}\hat{c}_{k}$, we have $%
\hat{V}_{nf}^{int}(-x)=\hat{c}_{nf}^{+int}(-x)\Lambda _{nf}^{int}(-x)+\hat{c}%
_{nf}^{int}(-x)\Lambda _{nf}^{+int}(-x)$ with $\Lambda
_{nf}^{int}(-x)=\sum_{k\in K_{n}}V_{k}^{(nf)}\hat{c}_{k}\exp (\frac{i}{\hbar
}\varepsilon _{k}x)$.

Similarly, writing the coupling Hamiltonian for energy transfer $\hat{W}%
=\sum_{n}\hat{W}_{n}$ as%
\begin{equation}
\hat{W}_{n}=b_{n}^{+}\Theta _{n}+b_{n}\Theta _{n}^{+}  \label{eq:V_Nn}
\end{equation}%
where $\Theta _{n}=\sum_{k\neq k^{\prime }\in K_{n}}W_{kk^{\prime
}}^{(n)}b_{k^{\prime }k}$, then

\begin{equation}
\hat{W}_{n}^{int}(-x)=b_{n}^{+int}(-x)\Theta
_{n}^{int}(-x)+b_{n}^{int}(-x)\Theta _{n}^{+int}(-x)  \label{eq:V_Nn^int}
\end{equation}%
where%
\begin{equation}
\Theta _{n}^{int}(-x)=\sum_{k\neq k^{\prime }\in K_{n}}W_{kk^{\prime
}}^{(n)}b_{k^{\prime }k}\exp [\frac{i}{\hbar }(\varepsilon _{k^{\prime
}}-\varepsilon _{k})x]  \label{eq:Teta_n_int}
\end{equation}

Bearing in mind that $\rho (t)=\sigma (t)\rho _{K}$ where $\sigma
(t)=Tr_{K}\rho (t)$ and Eqs.(\ref{eq:V_Mnf}), (\ref{eq:V_Nn}) and (\ref%
{eq:V_Nn^int}), we get for the second term on the RHS of Eq.(\ref%
{eq:density_matrix})
\begin{align}
& -\frac{1}{\hbar ^{2}}Tr_{K}\{\int_{0}^{\infty }dx[\hat{V},[\hat{V}%
^{int}(-x),\rho (t)]]\}=  \notag \\
& -\frac{1}{\hbar ^{2}}\int_{0}^{\infty }dx\{Tr_{K}[\hat{V}\hat{V}%
^{int}(-x)\rho _{K}]\sigma (t)  \notag \\
& -Tr_{K}[\hat{V}\rho _{K}\sigma (t)\hat{V}^{int}(-x)]-Tr_{K}[\hat{V}%
^{int}(-x)\rho _{K}\sigma (t)\hat{V}]  \notag \\
& +Tr_{K}[\rho _{K}\sigma (t)\hat{V}^{int}(-x)\hat{V}]\}  \label{eq:Tr1}
\end{align}
In evaluating the RHS of Eq.(\ref{eq:Tr1}) we encounter reservoir
correlation functions that reflect the reservoir equilibrium properties as
well as the nature of its interaction with the wire. For example,
\begin{align}
& C_{nf}(-x)=Tr_{K}[\Lambda _{nf}\Lambda _{nf}^{+}(-x)\rho _{K_{n}}]  \notag
\\
& =\sum_{k\in K_{n}}|V_{k}^{(nf)}|^{2}[1-f_{K_{n}}(\varepsilon _{k})]\exp (-%
\frac{i}{\hbar }\varepsilon _{k}x)  \label{eq:LambdaCorFunct}
\end{align}
Turning to the energy transfer contribution, third term on the RHS of Eq.(%
\ref{eq:density_matrix}), we obtain an expression of the form (\ref{eq:Tr1})
with the energy transfer interaction $\hat{W}$ replacing $\hat{V}$. Using
the Wick's theorem, we obtain correlation functions of the type

\begin{align}
& D_{n}(-x)=Tr_{K}[\Theta _{n}\Theta _{n}^{+}(-x)\rho _{K_{n}}]  \notag \\
& =\sum_{k\neq k^{\prime }\in K_{n}}\left\vert W_{kk^{\prime
}}^{(n)}\right\vert ^{2}f_{K_{n}}(\varepsilon _{k})[1-f_{K_{n}}(\varepsilon
_{k^{\prime }})]\exp [\frac{i}{\hbar }(\varepsilon _{k}-\varepsilon
_{k^{\prime }})x]  \label{eq:Teta_nTeta_n_int_1}
\end{align}%
Below we get a Markovian master equation in the wide-band limit. The full
master equation obtained in this way constitutes a set of 256 coupled
equation for the 16x16 elements of the wire density matrix, which can be
solved numerically by diagonalizing the corresponding Liouvillian matrix. In
particular we are interested in the steady state solution, $\sigma _{SS}$,
which is given by the eigenvector of zero eigenvalue. Once $\sigma _{SS}$
has been found, the current is obtained from%
\begin{equation}
\langle I\rangle =Tr(\hat{I}\sigma _{SS})  \label{eq:<current>}
\end{equation}%
where the current operator (defined, e.g., as the rate of change of electron
population on the left of the dashed line in Fig. \ref{fig:dimer_model}) is
given by

\begin{equation}
\hat{I}=e\frac{d}{dt}\hat{N}=\frac{ie}{\hbar }[\hat{H},\hat{N}]
\label{eq:current1}
\end{equation}
\begin{equation}
\hat{N}=\sum_{k\in L}\hat{c}_{k}^{+}\hat{c}_{k}+\hat{n}_{1g}+\hat{n}_{1e}
\label{eq:N}
\end{equation}%
In section \ref{sec:Numerical simulations} we show some results of such
numerical calculations. To gain better insight of the transport properties
of this model, analytical simplifications in some limits are useful. These
are discussed next.

\section{Analytical evaluation}

\label{sec:Analytical}

\begin{figure}[tbp]
\begin{center}
\includegraphics[width=7.cm,clip,angle=0]
{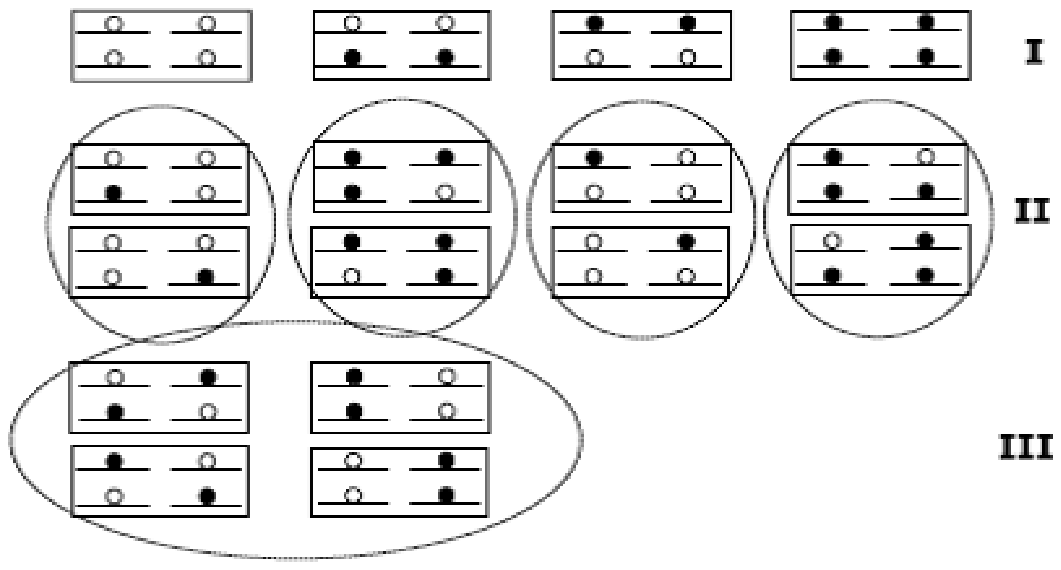}
\end{center}
\caption{A schematic display of the block structure of the wire Hamiltonian.}
\label{fig:subspaces}
\end{figure}

It is known \cite{Kaiser06} that for the evaluation of Eqs. (\ref%
{eq:density_matrix}) and (\ref{eq:Tr1}) it is essential to work in the
representation of the eigenstates of the Hamiltonian $\hat{H}_{wire}+\hat{H}%
_{leads}$ that defines the zeroth-order time evolution. The use of other
representations bears the danger of generating artifacts, which, for
instance, may lead to a violation of fundamental equilibrium properties \cite%
{Novotny02}. We thus face the problem of diagonalizing a matrix of order
256. This procedure may be facilitated by using the pseudospin description
based on the symmetry properties of Lie group SU(2) associated with the two
state problem ($1f,2f$); $f=e,g$. Such a \textquotedblleft donor
acceptor\textquotedblright\ system may be described by the \textquotedblleft
charge transfer\textquotedblright\ operators $b_{f}^{+}=\hat{c}_{2f}^{+}\hat{%
c}_{1f}$ and $b_{f}=\hat{c}_{1f}^{+}\hat{c}_{2f}$ that describe intersite
charge transfer $1\rightarrow 2$ and $2\rightarrow 1$, respectively, in
upper and lower states of the molecular dimer. The non-diagonal part of $%
\hat{H}_{wire}$, Eq.(\ref{eq:H_wire}), can then be written in terms of
operators $b_{f}$ only

\begin{equation}
\hat{H}_{wire}^{(nondiag)}=-\sum_{f=g,e}\Delta _{f}(b_{f}^{+}+b_{f})-\hbar
J(b_{e}^{+}b_{g}+b_{g}^{+}b_{e})  \label{eq:H_wire_up}
\end{equation}%
Define also the pseudospin (Bloch) vector in the second quantization picture

\begin{equation}
\left(
\begin{array}{c}
r_{1}^{f} \\
r_{2}^{f} \\
r_{3}^{f}%
\end{array}%
\right) =\left(
\begin{array}{c}
b_{f}^{+}+b_{f} \\
i(b_{f}-b_{f}^{+}) \\
\hat{n}_{2f}-\hat{n}_{1f}%
\end{array}%
\right) ;\text{ }f=g,e  \label{eq:uvw_M}
\end{equation}%
Its components have the following properties: (a) They satisfy the same
commutation rules as Pauli matrices $\hat{\sigma}_{1,2,3}$ \cite%
{All75,Hio81,yang-etal.04}; (b) the operators $\lambda _{f}=\hat{n}_{2f}+%
\hat{n}_{1f}=\sum_{m=1,2}\hat{c}_{mf}^{+}\hat{c}_{mf}$, $f=e,g$ (cf. Eq.(\ref%
{eq:lambda_f})) and $r_{i}^{f}$ commute: $[r_{i}^{f},\lambda _{f}]=0$ ($%
i=1,2,3$); (c) any linear operator of the "donor acceptor" system can be
written as linear superposition of the operators \{$r_{i}^{f}$\} and $%
\lambda _{f}$. In particular, the wire Hamiltonian can be written as
\begin{align}
\hat{H}_{wire}& =\frac{1}{2}\lambda _{e}(\varepsilon _{1e}+\varepsilon
_{2e})+\sum_{f=g,e}[\frac{1}{2}r_{3}^{f}(\varepsilon _{2f}-\varepsilon
_{1f})-\Delta _{f}r_{1}^{f}]  \notag \\
& -\frac{\hbar J}{2}(r_{1}^{e}r_{1}^{g}+r_{2}^{e}r_{2}^{g})+%
\sum_{m=1,2}U_{m}N_{m}(N_{m}-1)  \label{eq:H_wire_upBloch1}
\end{align}%
In Eq.(\ref{eq:H_wire_upBloch1}) we have put, without loss of generality, $%
(\varepsilon _{1g}+\varepsilon _{2g})/2=0$. Because the operators $\lambda
_{f}$ and $r_{i}^{f}$ commute, $\lambda _{f}$ is conserved under unitary
transformations related to the diagonalization of $\hat{H}_{wire}$.
Therefore, a total $2^{4}\times 2^{4}$ space can be partitioned into nine
smaller subspaces, i.e. the Liouvillian matrix in the required basis is
block diagonal with blocks, according to the values of $\lambda _{f}=0,1,2$
(see Fig.\ref{fig:subspaces}): four one-dimensional subspaces for $\lambda
_{f}=0,2$ for either $f=e,g$ (type I); four two-dimensional subspaces for $%
\lambda _{f}=1$ and $\lambda _{f^{\prime }}=0$,$2$ where $f\neq f^{\prime }$
(type II)$;$ and one four-dimensional subspace for $\lambda _{e}=\lambda
_{g}=1$ (type III). The type I submatrix is diagonal, while four state pairs
with each pair coupled by the charge transfer interaction are associated
with the four $2\times 2$ blocks of the type II subspace. The remaining four
states are coupled by both the charge transfer and exciton transfer
interaction and constitute the $4\times 4$ block of subspace III. Each of
these subspaces is characterized by assigning the values ($\lambda
_{e},\lambda _{g}$) of total populations in the ground and excited states of
the two bridge sites.

Using the identity

\begin{equation}
(r_{1}^{f})^{2}=(r_{2}^{f})^{2}=(r_{3}^{f})^{2}=\lambda _{f}-2\hat{n}_{2f}%
\hat{n}_{1f}=\left\{
\begin{array}{c}
0\text{ for }\lambda _{f}=0,2 \\
1\text{ for }\lambda _{f}=1%
\end{array}%
\right\} ,  \label{eq:norm1}
\end{equation}%
the wire Hamiltonian (\ref{eq:H_wire_upBloch1}) can be written in the form
\begin{align}
\hat{H}_{wire}& =\frac{1}{2}\lambda _{e}(\varepsilon _{1e}+\varepsilon
_{2e})+\sum_{m=1,2}U_{m}N_{m}(N_{m}-1)+  \notag \\
&
\begin{array}{c}
+0\text{ \ \ \ \ \ \ \ \ \ \ \ \ \ \ \ \ \ \ \ \ \ \ \ \ \ \ \ \ \ \ \ \ \ \
For subspaces I} \\
+\frac{1}{2}r_{3}^{f}(\varepsilon _{2f}-\varepsilon _{1f})-\Delta
_{f}r_{1}^{f}\text{ \ \ For subspaces II} \\
+[\frac{1}{2}\sum_{f=g,e}r_{3}^{f}(\varepsilon _{2f}-\varepsilon
_{1f})-\sum_{f=g,e}\Delta _{f}r_{1}^{f}- \\
-\frac{\hbar J}{2}(r_{1}^{e}r_{1}^{g}+r_{2}^{e}r_{2}^{g})]\text{ \ \ \ \ \ \
\ \ \ \ \ For subspace III}%
\end{array}
\label{eq:H_wire_upBloch_lambda_f1}
\end{align}

This prediagonalization provides an important simplification of our problem.
From Eqs. (\ref{eq:<current>}), (\ref{eq:current1}), (\ref{eq:N}) the
current is given by
\begin{equation}
\hat{I}=\frac{ie}{\hbar }\sum_{f=g,e}\Delta _{f}(b_{f}-b_{f}^{+})=\frac{e}{%
\hbar }\sum_{f=g,e}\Delta _{f}r_{2}^{f}  \label{eq:current operator}
\end{equation}
Using Eq.(\ref{eq:norm1}), this yields%
\begin{equation}
\hat{I}=\frac{e}{\hbar }\sum_{f=g,e}\Delta _{f}r_{2}^{f}(\lambda _{f}=1)
\label{eq:current_lambda=1}
\end{equation}

Obviously $\lambda _{f}=1$ in Eq.(\ref{eq:current_lambda=1}) is another way
of saying that the current in channel $\ f$ exists only for the case of one
of states $\{f\}$ is occupied and another one of $\{f\}$ is unoccupied.

Further simplification is made below, when we consider two specific limiting
cases. The first limit, $U_{m}=0$, describes noninteracting electrons at
each sites. In the opposite limit with strong on-site Coulomb repulsion $%
U_{m}$ $(m=1,2)$ is much larger than any other energy scale of the problem.
In the latter case we disregards states with more than one electron on any
of the molecular site $1$ and $2$ so we need to consider only 9 bridge
states: $|0_{1g},0_{2g},0_{1e},0_{2e}\rangle $, $%
|0_{1g},0_{2g},1_{1e},1_{2e}\rangle $ and $|1_{1g},1_{2g},0_{1e},0_{2e}%
\rangle $ in subspaces I; $|0_{1g},0_{2g},0_{1e},1_{2e}\rangle $, $%
|0_{1g},0_{2g},1_{1e},0_{2e}\rangle $, $|0_{1g},1_{2g},0_{1e},0_{2e}\rangle $
and $|1_{1g},0_{2g},0_{1e},0_{2e}\rangle $ in subspaces II; $%
|1_{1g},0_{2g},0_{1e},1_{2e}\rangle $ and $|0_{1g},1_{2g},1_{1e},0_{2e}%
\rangle $ in subspace III.

The diagonalization procedure yields the transformation between the
eigenstates of the wire Hamiltonian and the states of the non-interaction
molecular wire, $|n_{1g},n_{2g},n_{1e},n_{2e}\rangle $ displayed in Fig.\ref%
{fig:subspaces}. Denoting the column vectors of these states by $\{\Phi \}$
and $\{\chi \}$, respectively, and the transformation between them by $\hat{Y%
}$, i.e. $\{\chi \}=\hat{Y}\{\Phi \}$, we can characterized each eigenstate $%
\Phi $ by the corresponding subspace $(\lambda _{e},\lambda _{g})$. In this
basis, the fermionic interaction picture operators (see Eq.(\ref%
{eq:interaction represent})) read, for example%
\begin{widetext}
\begin{align*}
& \langle \alpha |\hat{c}_{nf}^{int}(-x)|\beta \rangle=
\lbrack \hat{Y}^{+}(\lambda _{e}(\alpha ),\lambda _{g}(\alpha ))\tilde{\chi%
}^{+}(\lambda _{e}(\alpha ),\lambda _{g}(\alpha ))\hat{c}_{nf}\tilde{\chi}%
(\lambda _{e}(\beta ),\lambda _{g}(\beta ))\hat{Y}(\lambda _{e}(\beta
),\lambda _{g}(\beta ))]_{\alpha \beta } \\
& \times \exp [\frac{i}{\hbar }(E_{\beta }(\lambda _{e}(\beta ),\lambda
_{g}(\beta ))-E_{\alpha }(\lambda _{e}(\alpha ),\lambda _{g}(\alpha )))x]
\end{align*}%
\end{widetext}
where $(\lambda _{e}(\alpha ),\lambda _{g}(\alpha ))$ denotes the subspace
associated with the eigenstate $\alpha $ and points to the corresponding
values of $\lambda _{e}$ and $\lambda _{g}$, and where $(\lambda _{e}(\beta
),\lambda _{g}(\beta ))=$ $(\lambda _{e}(\alpha )+1,\lambda _{g}(\alpha ))$
if $f=e$ and $(\lambda _{e}(\beta ),\lambda _{g}(\beta ))=$ $(\lambda
_{e}(\alpha ),\lambda _{g}(\alpha )+1)$ if $f=g$. $\tilde{\chi}$ denotes the
transpose matrix $\hat{\chi}$. The relaxation terms in the master equation (%
\ref{eq:density_matrix}) take in this basis the forms
\begin{widetext}
\begin{align}
& -\frac{1}{\hbar ^{2}}Tr_{K}\int_{0}^{\infty }dx[\hat{V},[\hat{V}%
^{int}(-x),\rho (t)]]_{\alpha \beta }  
 =\frac{1}{2}\sum_{nf\alpha ^{\prime }\beta ^{\prime }}\Gamma _{nf}\{\hat{c}%
_{nf,\alpha \alpha ^{\prime }}\sigma _{\alpha ^{\prime }\beta ^{\prime }}%
\hat{c}_{nf,\beta ^{\prime }\beta }^{+}[2-f_{K_{n}}(E_{\beta ^{\prime
}}-E_{\beta })  \notag \\
& -f_{K_{n}}(E_{\alpha ^{\prime }}-E_{\alpha })]+\hat{c}_{nf,\alpha \alpha
^{\prime }}^{+}\sigma _{\alpha ^{\prime }\beta ^{\prime }}\hat{c}_{nf,\beta
^{\prime }\beta }[f_{K_{n}}(E_{\beta }-E_{\beta ^{\prime }}) 
 +f_{K_{n}}(E_{\alpha }-E_{\alpha ^{\prime }})]-\{\hat{c}_{nf,\alpha \alpha
^{\prime }}\hat{c}_{nf,\alpha ^{\prime }\beta ^{\prime
}}^{+}f_{K_{n}}(E_{\alpha ^{\prime }}-E_{\beta ^{\prime }})  \notag \\
& +\hat{c}_{nf,\alpha \alpha ^{\prime }}^{+}\hat{c}_{nf,\alpha ^{\prime
}\beta ^{\prime }}[1-f_{K_{n}}(E_{\beta ^{\prime }}-E_{\alpha ^{\prime
}})]\}\sigma _{\beta ^{\prime }\beta }  
 -\sigma _{\alpha \alpha ^{\prime }}\{\hat{c}_{nf,\alpha ^{\prime }\beta
^{\prime }}\hat{c}_{nf,\beta ^{\prime }\beta }^{+}f_{K_{n}}(E_{\beta
^{\prime }}-E_{\alpha ^{\prime }})  
 +\hat{c}_{nf,\alpha ^{\prime }\beta ^{\prime }}^{+}\hat{c}_{nf,\beta
^{\prime }\beta }[1-f_{K_{n}}(E_{\alpha ^{\prime }}-E_{\beta ^{\prime
}})]\}\}  \label{eq:sigma_albe}
\end{align}%
\end{widetext}

where

\begin{equation}
\Gamma _{nf}=\frac{2\pi }{\hbar }\sum_{k\in K_{n}}|V_{k}^{(nf)}|^{2}\delta
(\varepsilon _{k}-\varepsilon _{nf})  \label{eq:Gamma}
\end{equation}%
and

\begin{widetext}
\begin{align}
& -\frac{1}{\hbar ^{2}}Tr_{K}\int_{0}^{\infty }dx[\hat{W},[\hat{W}%
^{int}(-x),\rho (t)]]_{\alpha \beta }  \notag \\
& =\frac{1}{2}\sum_{n\alpha ^{\prime }\beta ^{\prime }}\{-B_{K_{n}}[E_{\beta
^{\prime }}(\lambda _{e}+1,\lambda _{g})-E_{\alpha ^{\prime }}(\lambda
_{e},\lambda _{g}+1),\mu _{K_{n}}][b_{n,\alpha \alpha ^{\prime
}}^{+}b_{n,\alpha ^{\prime }\beta ^{\prime }}\sigma _{\beta ^{\prime }\beta
}(t)+\sigma _{\alpha \beta ^{\prime }}(t)b_{n,\beta ^{\prime }\alpha
^{\prime }}^{+}b_{n,\alpha ^{\prime }\beta }]  \notag \\
& -B_{K_{n}}[E_{\alpha ^{\prime }}(\lambda _{e},\lambda _{g}+1)-E_{\beta
^{\prime }}(\lambda _{e}+1,\lambda _{g}),\mu _{K_{n}}][b_{n,\alpha \beta
^{\prime }}b_{n,\beta ^{\prime }\alpha ^{\prime }}^{+}\sigma _{\alpha
^{\prime }\beta }(t)+\sigma _{\alpha \alpha ^{\prime }}(t)b_{n,\alpha
^{\prime }\beta ^{\prime }}b_{n,\beta ^{\prime }\beta }^{+}]  \notag \\
& +b_{n,\alpha \alpha ^{\prime }}^{+}\sigma _{\alpha ^{\prime }\beta
^{\prime }}(t)b_{n,\beta ^{\prime }\beta }B_{K_{n}}[E_{\beta ^{\prime
}}(\lambda _{e},\lambda _{g}+1)-E_{\beta }(\lambda _{e}+1,\lambda _{g}),\mu
_{K_{n}}]  \notag \\
& +b_{n,\alpha \alpha ^{\prime }}\sigma _{\alpha ^{\prime }\beta ^{\prime
}}(t)b_{n,\beta ^{\prime }\beta }^{+}B_{K_{n}}[E_{\beta ^{\prime }}(\lambda
_{e}+1,\lambda _{g})-E_{\beta }(\lambda _{e},\lambda _{g}+1),\mu _{K_{n}}]
\notag \\
& +b_{n,\alpha \alpha ^{\prime }}^{+}\sigma _{\alpha ^{\prime }\beta
^{\prime }}(t)b_{n,\beta ^{\prime }\beta }B_{K_{n}}[E_{\alpha ^{\prime
}}(\lambda _{e},\lambda _{g}+1)-E_{\alpha }(\lambda _{e}+1,\lambda _{g}),\mu
_{K_{n}}]  \notag \\
& +b_{n,\alpha \alpha ^{\prime }}\sigma _{\alpha ^{\prime }\beta ^{\prime
}}(t)b_{n,\beta ^{\prime }\beta }^{+}B_{K_{n}}[E_{\alpha ^{\prime }}(\lambda
_{e}+1,\lambda _{g})-E_{\alpha }(\lambda _{e},\lambda _{g}+1),\mu _{K_{n}}]\}
\label{eq:V_Nterm2}
\end{align}%
\end{widetext}
where%
\begin{widetext}
\begin{equation}
B_{K_{n}}(E_{\alpha }-E_{\beta },\mu _{K_{n}})=\frac{2\pi }{\hbar }%
\sum_{k\neq k^{\prime }\in K_{n}}\left\vert W_{kk^{\prime
}}^{(n)}\right\vert ^{2}\delta (\varepsilon _{k}-\varepsilon _{k^{\prime
}}+E_{\alpha }-E_{\beta })f_{K_{n}}(\varepsilon
_{k})[1-f_{K_{n}}(\varepsilon _{k^{\prime }})]  \label{eq:B_NKn}
\end{equation}
\end{widetext}

In evaluating these forms we have taken the wide band limit for the
electrodes spectral densities.

Next consider the diagonalization procedure itself. In subspaces I the
unitary transformation $\hat{Y}(\lambda _{e},\lambda _{g})$ is obviously the
unity matrix. The diagonalization of the block matrices in subspaces II and
III is carry out in the limiting cases of zero and infinite on-site
interactions.

\subsection{Zero on site coupling}

The case of zero on site coupling is discussed in Appendix A. We find the
eigenfunctions and energies of the 2-site bridge summarized in Table I

\begin{widetext}
$%
\begin{array}
[c]{cccc}
& \lambda_{g}=0 & \lambda_{g}=1 & \lambda_{g}=2\\
\lambda_{e}=0 &
\begin{array}
[c]{c}%
\Phi(0,0)=\\
=|0_{1g},0_{2g},0_{1e},0_{2e}\rangle\\
E=0
\end{array}
&
\begin{array}
[c]{c}%
\begin{array}
[c]{c}%
\Phi_{+}(0,1)=|0_{1g},1_{2g},0_{1e},0_{2e}\rangle\\
\Phi_{-}(0,1)=|1_{1g},0_{2g},0_{1e},0_{2e}\rangle
\end{array}
\\
E=0
\end{array}
&
\begin{array}
[c]{c}%
\Phi(0,2)=|1_{g},1_{2g},0_{e},0_{2e}\rangle\\
E=0
\end{array}
\\
\lambda_{e}=1 &
\begin{array}
[c]{c}%
\begin{array}
[c]{c}%
\Phi_{\pm}(1,0)=\\
=\frac{1}{\sqrt{2}}(|0_{1g},0_{2g},0_{1e},1_{2e}\rangle\mp\\
\mp|0_{1g},0_{2g},1_{1e},0_{2e}\rangle)\\
E=\varepsilon_{2e}\pm\Delta_{e}%
\end{array}
\\
\hat{\chi}(1,0)=\left(
\begin{array}
[c]{c}%
|0_{1g},0_{2g},0_{1e},1_{2e}\rangle\\
|0_{1g},0_{2g},1_{1e},0_{2e}\rangle
\end{array}
\right)
\end{array}
&
\begin{array}
[c]{c}%
\begin{array}
[c]{c}%
\hat{\Phi}(1,1)=Y^{+}(1,1)\hat{\chi}(1,1)\\
E=\frac{1}{2}(\varepsilon_{1e}+\varepsilon_{2e})\pm\frac{1}{2}J\hbar\pm\\
\pm\frac{1}{2}\sqrt{4\Delta_{e}^{2}+J^{2}\hbar^{2}}%
\end{array}
\\
\hat{\chi}(1,1)=\left(
\begin{array}
[c]{c}%
|1_{1g},0_{2g},1_{1e},0_{2e}\rangle\\
|1_{1g},0_{2g},0_{1e},1_{2e}\rangle\\
|0_{1g},1_{2g},1_{1e},0_{2e}\rangle\\
|0_{1g},1_{2g},0_{1e},1_{2e}\rangle
\end{array}
\right)
\end{array}
&
\begin{array}
[c]{c}%
\begin{array}
[c]{c}%
\Phi_{\pm}(1,2)=\\
=\frac{1}{\sqrt{2}}(|1_{1g},1_{g},0_{1e},1_{e}\rangle\mp\\
\mp|1_{1g},1_{2g},1_{1e},0_{2e}\rangle)\\
E=\varepsilon_{2e}\pm\Delta_{e}%
\end{array}
\\
\hat{\chi}(1,2)=\left(
\begin{array}
[c]{c}%
|1_{g},1_{2g},0_{e},1_{2e}\rangle\\
|1_{g},1_{2g},1_{e},0_{2e}\rangle
\end{array}
\right)
\end{array}
\\
\lambda_{e}=2 &
\begin{array}
[c]{c}%
\Phi(2,0)=|0_{1g},0_{2g},1_{1e},1_{2e}\rangle\\
E=\varepsilon_{1e}+\varepsilon_{2e}%
\end{array}
&
\begin{array}
[c]{c}%
\begin{array}
[c]{c}%
\Phi_{+}(2,1)=|0_{1g},1_{2g},1_{1e},1_{2e}\rangle\\
\Phi_{-}(2,1)=|1_{1g},0_{2g},1_{1e},1_{2e}\rangle
\end{array}
\\
E=\varepsilon_{1e}+\varepsilon_{2e}%
\end{array}
&
\begin{array}
[c]{c}%
\Phi(2,2)=|1_{1g},1_{2g},1_{1e},1_{2e}\rangle\\
E=\varepsilon_{1e}+\varepsilon_{2e}%
\end{array}
\end{array}
$
\end{widetext}

where%
\begin{equation}
Y^{+}(1,1)=\frac{1}{\sqrt{2}}\left(
\begin{array}{cccc}
\sin \tau & -\cos \tau & -\cos \tau & \sin \tau \\
\cos \tau & \sin \tau & \sin \tau & \cos \tau \\
\sin \tau & \cos \tau & -\cos \tau & -\sin \tau \\
\cos \tau & -\sin \tau & \sin \tau & -\cos \tau%
\end{array}%
\right) ,  \label{eq:Y(1,1)}
\end{equation}%
and where $\tau $ is given by
\begin{equation}
\cos 2\tau =\frac{-J\hbar }{\sqrt{4\Delta _{e}^{2}+J^{2}\hbar ^{2}}}\text{
and \ \ }\sin 2\tau =\frac{2\Delta _{e}}{\sqrt{4\Delta _{e}^{2}+J^{2}\hbar
^{2}}}  \label{eq:cos_sin}
\end{equation}%
The current in this case is found to be

\begin{align}
\langle I\rangle & =-\frac{2e}{\hbar }\Delta _{e}\mathrm{Im}\{[\sigma
_{32}(1,1)+\sigma _{41}(1,1)]\cos 2\tau  \notag \\
& +[\sigma _{31}(1,1)-\allowbreak \sigma _{42}(1,1)]\sin 2\tau
-\sum_{\lambda _{g}=0,2}\sigma _{-+}(1,\lambda _{g})\}
\label{eq:current_final1}
\end{align}%
Indices "$+$" and "$-$" in Eq.(\ref{eq:current_final1}) correspond to the
functions $\Phi _{+}(1,\lambda _{g})$ and $\Phi _{-}(1,\lambda _{g})$,
respectively, in Table I. Indices $1,2,3$ and $4$ label the the eigenstates
of the wire Hamiltonian in subspace III. The corresponding energies are
given by formulas $E_{1}\equiv E_{-,+}$, $E_{2}\equiv E_{-,-},$ $E_{3}\equiv
E_{+,-},$ $E_{4}\equiv E_{+,+}$ where
\begin{equation}
E_{\pm ,\pm }=\varepsilon _{e}\pm \frac{1}{2}J\hbar \pm \frac{1}{2}\sqrt{%
4\Delta _{e}^{2}+J^{2}\hbar ^{2}}  \label{eq:E+-+-}
\end{equation}

\subsection{Rotating-wave approximation}

\label{sec:RWA}

The calculation of the non-diagonal elements of the density matrix $\sigma
_{\alpha \beta }(1,\lambda _{g})$ in Eq.(\ref{eq:current_final1}) for the
current is essentially simplified for very weak wire--lead coupling when the
coherent time-evolution dominates the dynamics of the wire electrons. This
means that the largest time-scale of the coherent evolution, given by the
smallest energy difference, and the dissipative time-scale, determined by
the electron and energy transfer rates, $\Gamma _{nf}$ and $%
B_{K_{n}}(E_{\alpha }-E_{\beta },\mu _{K_{n}})$, respectively, are well
separated, i.e., $\hbar \Gamma _{nf},\hbar B_{K_{n}}(E_{\alpha }-E_{\beta
},\mu _{K_{n}})\ll \left\vert E_{\alpha }-E_{\beta }\right\vert $ for $%
\lambda _{e}=1$ and $\alpha \neq \beta $. Then for $\lambda _{e}=1$ and $%
\alpha \neq \beta $, Eq.(\ref{eq:density_matrix}) is dominated by the first
term on the RHS. Consequently, $\sigma _{\alpha \beta }(1,\lambda _{g})$ can
be calculated in the first order of $\hbar \Gamma _{nf}/(E_{\alpha
}-E_{\beta })$ and $\hbar B_{K_{n}}(E_{\alpha }-E_{\beta },\mu
_{K_{n}})/(E_{\alpha }-E_{\beta })$. This constitutes the essence of a
rotating-wave approximation (RWA) \cite{Kaiser06}. Within it, one can
provide a closed expression for the reduced density matrix elements $\sigma
_{\alpha \beta }$ and for the stationary current. We shall use the RWA below
in Sec.\ref{subsec:exciton_induced_current} and Appendix B.

\subsection{Strong Coulomb repulsion at sites}

\label{sec:Strong Coulomb repulsion at sites}

In the limit of strong Coulomb repulsion, $U_{m}$ is assumed to be so large
that at most one excess electron resides on each site. Thus, the available
Hilbert space for uncoupled sites is reduced to three states $\hat{\chi}%
(0,0)=|0_{1g},0_{2g},0_{1e},0_{2e}\rangle $, $\hat{\chi}%
(2,0)=|0_{1g},0_{2g},1_{1e},1_{2e}\rangle $ and $\hat{\chi}%
(0,2)=|1_{1g},1_{2g},0_{1e},0_{2e}\rangle $ for subspaces I; two states $%
\hat{\chi}(1,0)=\left(
\begin{array}{c}
|0_{1g},0_{2g},0_{1e},1_{2e}\rangle \\
|0_{1g},0_{2g},1_{1e},0_{2e}\rangle%
\end{array}%
\right) $ and $\hat{\chi}(0,1)=\left(
\begin{array}{c}
|0_{1g},1_{2g},0_{1e},0_{2e}\rangle \\
|1_{1g},0_{2g},0_{1e},0_{2e}\rangle%
\end{array}%
\right) $ for subspaces II; and the state $\hat{\chi}(1,1)=\left(
\begin{array}{c}
|1_{1g},0_{2g},0_{1e},1_{2e}\rangle \\
|0_{1g},1_{2g},1_{1e},0_{2e}\rangle%
\end{array}%
\right) $ for the now 2-dimensional subspace III. The unitary operators $%
\hat{Y}^{+}(1,0)$ and $\hat{Y}^{+}(0,1)$ and the corresponding eigenstates
and eigenvalues are defined by the same Eqs. (\ref{eq:Psi_f+-single}), (\ref%
{eq:Psi_f+-1}) and (\ref{eq:eigenvalues_noncollective1}), respectively, as
before (see Appendix A). The operator $\hat{Y}^{+}(1,1)$ is reduced to (see
Appendix B)

\begin{equation}
\hat{Y}^{+}(1,1)=\frac{1}{\sqrt{2}}\left(
\begin{array}{cc}
1 & 1 \\
-1 & 1%
\end{array}%
\right)  \label{eq:Y+(1,1)blockade}
\end{equation}%
$\hat{Y}^{+}(1,1)$ is used to obtain the corresponding eigenstates $\hat{\Phi%
}(1,1)=Y^{+}(1,1)\hat{\chi}(1,1)$ and eigenvalues $E_{1,2}=\varepsilon
_{e}\mp J\hbar $.

Substituting Eq.(\ref{eq:Y+(1,1)blockade}) into Eq.(\ref{eq:current}) of
Appendix A for the current, we get for $\Delta _{g}=0$%
\begin{equation}
\langle I\rangle =\frac{2e}{\hbar }\Delta _{e}\mathrm{Im}\sigma _{-+}(1,0)
\label{eq:currentCB}
\end{equation}

\section{Current from the energy transfer interaction in the wire}

\label{subsec:exciton_induced_current}

In a recent paper \cite{Nitzan06PRL} Galperin, Nitzan and Ratner have
predicted the existence of non-Landauer current induced by energy transfer
interactions between a bridge molecule and electron-hole excitations in the
leads. Here we show that a similar non-Landauer current arises from the
exciton type interaction $J$ in the wire itself. For simplicity we limit
ourselves to electron transfer interaction between the wire and the metal
leads, Eq.(\ref{eq:sigma_albe}), and disregard the corresponding excitation
transfer, Eq.(\ref{eq:V_Nterm2}). Also for simplicity we consider a large
bias limit in the Coulomb blockade case when $\mu _{L}>\varepsilon _{e}$ and
$\mu _{R}<\varepsilon _{g},$ and the states $\varepsilon _{e},\varepsilon
_{g}$ are positioned rather far ($\gg k_{B}T,\hbar |J|,|\Delta _{e}|$) from
the Fermi levels of both leads so that $f_{L}(\varepsilon )=1$ and $%
f_{R}(\varepsilon )=0$ can be taken on the RHS of Eq.(\ref{eq:sigma_albe}).
Finally, we disregard electron transfer interaction in the "$g$" channel,
i.e. we take $\Delta _{g}=0$. Landauer type current would be realized in
channel "$e$" when it is isolated from channel "$g$", i.e. when $J=0,$ $%
\Gamma _{1g}=\Gamma _{2g}=0$ and $\lambda _{g}=0$. Solving Eqs. (\ref%
{eq:density_matrix}), (\ref{eq:sigma_albe}) in the RWA approximation under
these conditions and substituting the steady-state solution into Eq.(\ref%
{eq:currentCB}), we get, using also the normalization condition $%
\sum_{\lambda _{e}=0,1,2}Tr\sigma (\lambda _{e},0)=1$,
\begin{equation}
\langle I\rangle _{RWA}=-e\frac{\Gamma _{1e}\Gamma _{2e}}{\Gamma
_{1e}+\Gamma _{2e}}  \label{eq:current_RWA}
\end{equation}%
Eq.(\ref{eq:current_RWA}) describes the Landauer current and coincides with
Eq.(21) of Ref.\cite{Kaiser06} (excluding the sign).

In fact, the current vanishes for $\Gamma _{1e}=\Gamma _{2e}=0$ even when $%
\Gamma _{1g},\Gamma _{2g}\neq 0$, since $\Delta _{g}=0$ (see Fig.\ref%
{fig:specific_coupling}). Such selective coupling to the leads could be
obtained for the bridge made of a quadruple quantum dot where the lateral
ones are strongly coupled to the leads \cite{Bryllert02,Brandes08PRB}.

\begin{figure}[tbp]
\begin{center}
\includegraphics[width=7.cm,clip,angle=0]
{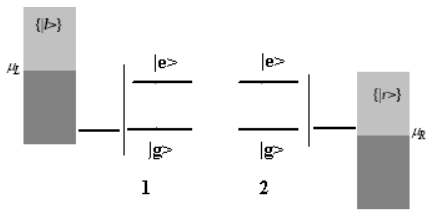}
\end{center}
\caption{A possible physical realization of the selective tunneling
configuration, where only $\left\vert g\right\rangle $ levels are coupled to
the leads. }
\label{fig:specific_coupling}
\end{figure}

\begin{figure}[tbph]
\begin{center}
\includegraphics[width=7.cm,clip,angle=0]
{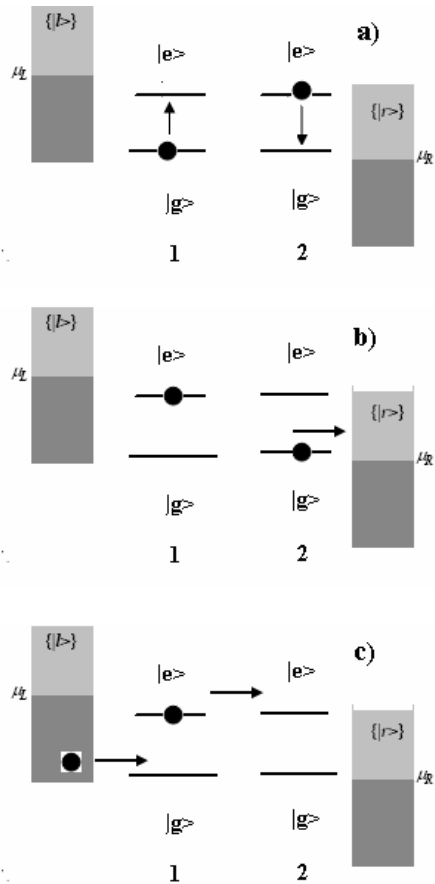}
\end{center}
\caption{Different stages of the energy-transfer induced current. a) energy
transfer, $\protect\sigma (1,1)\neq 0$. b) the charge transfer to the right
lead. c) the intersite charge transfer, $\protect\sigma (1,0)\neq 0$; the
charge transfer from the left lead.}
\label{fig:nonlan}
\end{figure}

Consider now the case when $\Gamma _{1e}=\Gamma _{2e}=0$; $\Gamma
_{1g},\Gamma _{2g}\neq 0$; $\Delta _{g}=0$ and $J\neq 0$. For this case Eqs.(%
\ref{eq:density_matrix}), (\ref{eq:sigma_albe}) together with Eq.(\ref%
{eq:currentCB}) lead to%
\begin{equation}
\langle I\rangle =-4e\Gamma _{g}J^{2}\Delta _{e}^{2}\frac{1-[\sigma
_{--}(0,1)+\sigma (2,0)]}{\Delta _{e}^{2}\Gamma _{g}{}^{2}+16\Delta
_{e}^{2}J^{2}+\hbar ^{2}\Gamma _{g}{}^{2}J^{2}}
\label{eq:nonLandauer_current_noRWA}
\end{equation}%
where for simplicity we put $\Gamma _{1g}=\Gamma _{2g}\equiv \Gamma _{g}$.
Eq.(\ref{eq:nonLandauer_current_noRWA}) describes a non-Landauer current
caused by transport in different channels: the intersite transfer occurs in
channel "$e$", and the charge transfer between the molecular bridge and the
leads occurs in channel "$g$". The interchannel mixing is induced by the
energy-transfer term $J$ (see Fig.\ref{fig:nonlan}). For example, starting
with the molecular system in state $|1_{1g},0_{2g},0_{1e},1_{2e}\rangle $,
electron transmission takes place along route such as $%
|1_{1g},0_{2g},0_{1e},1_{2e}\rangle \overset{1}{\longrightarrow }%
|0_{1g},1_{2g},1_{1e},0_{2e}\rangle \overset{2}{\longrightarrow }%
|0_{1g},0_{2g},1_{1e},0_{2e}\rangle \overset{3}{\longrightarrow }%
|0_{1g},0_{2g},0_{1e},1_{2e}\rangle \overset{4}{\longrightarrow }%
|1_{1g},0_{2g},0_{1e},1_{2e}\rangle $. Step 1 is an energy transfer process,
steps 2 and 3 rely on $\Gamma _{2g}\neq 0$ and $\Delta _{e}\neq 0$,
respectively, and step 4 closes the circle via the $\Gamma _{1g}$ process.

Eq.(\ref{eq:nonLandauer_current_noRWA}) clearly shows that the current
exists only for $J\neq 0$ and $\Delta _{e}\neq 0.$ For small $J$, $\langle
I\rangle \sim J^{2}$. For large $J$ we obtain
\begin{equation}
\langle I\rangle \simeq -4e\Gamma _{g}\Delta _{e}^{2}\frac{1-[\sigma
_{--}(0,1)+\sigma (2,0)]}{16\Delta _{e}^{2}+\hbar ^{2}\Gamma _{g}{}^{2}}
\label{eq:currentCBstrongJ}
\end{equation}%
which does not depend on $J$. In the limit $\hbar \left\vert J\right\vert
,\left\vert \Delta _{e}\right\vert \gg \Gamma _{g}$, Eq.(\ref%
{eq:nonLandauer_current_noRWA}) yields for $\sigma _{--}(0,1)=\sigma (2,0)=0$

\begin{equation}
\langle I\rangle _{RWA}=-\frac{e}{2}\frac{\Gamma _{2g}\Gamma _{1g}}{\Gamma
_{2g}+\Gamma _{1g}}  \label{eq:nonLandauer_current}
\end{equation}%
In deriving Eq.(\ref{eq:nonLandauer_current}) we have not put $\Gamma
_{1g}=\Gamma _{2g}$. This limit corresponds to the range of validity of the
RWA. Indeed, it can be shown that Eq. (\ref{eq:nonLandauer_current}) can be
obtained for this model in the RWA (see Appendix C).

If $\sigma _{--}(0,1),\sigma (2,0)\neq 0,$ the non-Landauer current
decreases, since the populations of states $|1_{1g},0_{2g},0_{1e},0_{2e}%
\rangle $ ($\sigma _{--}(0,1)$) and $|0_{1g},0_{2g},1_{1e},1_{2e}\rangle $ ($%
\sigma (2,0)$) suppress current. Two latter states are also steady-states in
the case under consideration (Coulomb \ blocking, $\Gamma _{1e}=\Gamma
_{2e}=\Delta _{g}=0$) along with the states described by Fig.\ref{fig:nonlan}%
. The existence of several steady-states corresponds to the presence of the
respective zero eigenvalues of the relaxation matrix. Our numerical
calculations give three such zero eigenvalues corresponding to three above
steady-states.

If $\Gamma _{1e},\Gamma _{2e}\neq 0,$ state $|1_{1g},0_{2g},0_{1e},0_{2e}%
\rangle $ is only steady-state that \textquotedblleft
locks\textquotedblright\ the current due to Coulomb \ blocking, since $%
\Delta _{g}=0$. Numerical simulations of other situations when $\Delta
_{g}\neq 0$ and non-interacting electrons at a site are carried out in the
next section.

\section{Numerical results}

\label{sec:Numerical simulations}

The results presented in this section are based on direct numerical solution
of Eq. (\ref{eq:density_matrix}), and are in complete agreement with the
analytical solutions when applied to the special cases treated in Sections %
\ref{sec:Analytical} and \ref{subsec:exciton_induced_current}. The numerical
solution was carried using the the basis of eigenstates of the Hamiltonian $%
\hat{H}_{wire}$, Eq.(\ref{eq:H_wire}). Once $\sigma (t)$ is obtained from
Eq. (\ref{eq:density_matrix}), the expectation value of the current is
calculated as $\langle I\rangle =Tr(\hat{I}\sigma (t))$ where the current
operator was defined by Eq.(\ref{eq:current_lambda=1}). In this calculation
we have limited ourselves to the case where the wire-leads energy transfer
coupling $\hat{W}$ is disregarded and, unless otherwise specified, have used
the following parameters: $\varepsilon _{1g}=\varepsilon _{2g}=0.0eV$, $%
\varepsilon _{1e}=\varepsilon _{2e}=2.0eV$, $\Delta _{g}=\Delta _{e}=0.01eV$%
, $\Gamma _{1f}=\Gamma _{2f}=0.02eV$ for $f=g,e$ (below we use $\Gamma $ to
denote the order of magnitude of these widths) and $T=100K$. The Fermi
levels were taken to align symmetrically with respect to the energy levels $%
\varepsilon _{1g}$ and $\varepsilon _{1e}$, i.e. $\mu _{L}=(\varepsilon
_{1g}+\varepsilon _{1e}+V_{bs})/{2}$ and $\mu _{R}=\mu _{L}-V_{bs}$. We also
used the value of $e\Delta _{e}/\hbar =2.45\bullet 10^{-6}A$ as the unit of
current $\langle I\rangle $.

\begin{figure}[tbp]
\begin{center}
\includegraphics[width=7.cm,clip,angle=0]
{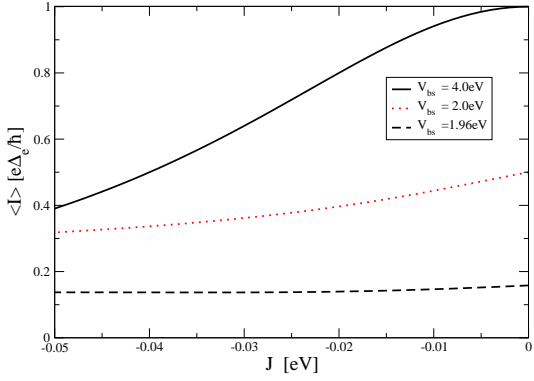}
\end{center}
\caption{The current $\langle I\rangle $ displayed as a function of the
exciton coupling parameter $J$ . $V_{bs}=1.96eV$ (dashed line); $%
V_{bs}=2.0eV $ (dotted line); $V_{bs}=4.0eV$ (solid line)}
\label{fig:b2040j}
\end{figure}

Consider first non-interacting electrons. Figs.\ref{fig:b2040j}, \ref%
{fig:b40jplus} and \ref{fig:b40jplusg0} show the expectation value of the
current $\langle I\rangle $ and one-particle populations $P_{nf}=Tr(\hat{c}%
_{nf}^{+}\hat{c}_{nf}\sigma )$ as functions of the exciton
interaction parameter $J$. One can see that if the imposed voltage bias $%
V_{bs}$ is larger than $\varepsilon _{e}-\varepsilon _{g}$, the expectation
value of the current diminishes when $\left\vert J\right\vert $ increases
(we have used $J<0$ which is typical to J-aggregates, however the trend is
similar with $J>0$). Such a behavior can be understood, using Eq.(\ref%
{eq:E+-+-}) for the energies in subspaces (III) and Eq.(\ref%
{eq:current_final1}) for the current. The latter equation shows two direct
contributions to the current. The first one has its origin in states of
subspace (III), the energies of which depend on both $\Delta _{e}$ and $J$
(the first and the second terms on the RHS of Eq.(\ref{eq:current_final1})).
The second contribution arises from states of subspaces (II), the energies
of which depend on $\Delta _{e}$ only (the third terms on the RHS of Eq.(\ref%
{eq:current_final1})). Using Eqs.(\ref{eq:density_matrix}) with $\hat{W}=0$
and (\ref{eq:sigma_albe}), the non-diagonal elements of the density matrix
on the RHS of Eq.(\ref{eq:current_final1}) for the steady-state condition
can be evaluated as $\sigma _{\alpha \beta }\sim \frac{-i\hbar \Gamma }{%
E_{\alpha }-E_{\beta }}$ (see also Sec.\ref{sec:RWA}). Since $%
E_{-}-E_{+}=-2\Delta _{e}$, Eq.(\ref{eq:eigenvalues_noncollective1}), we get
for the contribution of the third term on the RHS of Eq.(\ref%
{eq:current_final1})

\begin{figure}[tbp]
\begin{center}
\includegraphics[width=7.cm,clip,angle=0]
{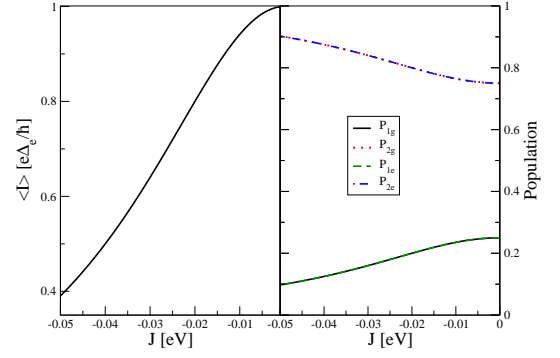}
\end{center}
\caption{The current $\langle I\rangle $ as a function of the parameter $J$
in the case of non- interacting electrons for $V_{bs}=4.0eV$. The current $%
\langle I\rangle $ is shown in left panel, and the populations $P_{1g}$, $%
P_{1e}$, $P_{2g}$ and $P_{2e}$ are shown in the right panel.}
\label{fig:b40jplus}
\end{figure}

\begin{figure}[tbp]
\begin{center}
\includegraphics[width=7.cm,clip,angle=0]
{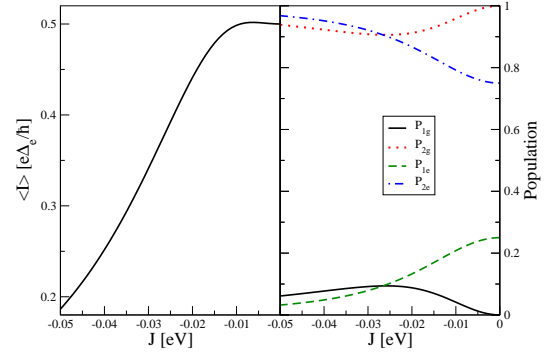}
\end{center}
\caption{Same as Fig.\protect\ref{fig:b40jplus} for the parameters $\Delta
_{g}=0,$ $\Delta _{e}=0.01eV$ and $V_{bs}=4.0eV$. }
\label{fig:b40jplusg0}
\end{figure}

\begin{equation}
\frac{2e}{\hbar }\Delta _{e}\mathrm{Im}\sum_{\lambda _{g}=0,2}\sigma
_{-+}(1,\lambda _{g})\sim 2e\Gamma  \label{eq:third}
\end{equation}%
The contribution of the first and the second terms on the RHS of Eq.(\ref%
{eq:current_final1}) depends on the relation between $J$ and $\Delta _{e}$.

When $\hbar \left\vert J\right\vert <<\Delta _{e}$, Eq.(\ref{eq:cos_sin})
yields $\cos 2\tau \approx 0$, $\sin 2\tau \approx 1$, and only the second
term in (\ref{eq:current_final1}) gives a contribution to the current from
the states of subspace (III). Under this condition one gets from Eq.(\ref%
{eq:E+-+-}) two doubly-degenerated values of energy $E_{1}=E_{4}=\varepsilon
_{e}+\Delta _{e}$ and $E_{2}=E_{3}=\varepsilon _{e}-\Delta _{e}$, where the
splitting is of the same order of magnitude as the hopping matrix element $%
\Delta _{e}$. We obtain
\begin{equation}
\frac{2e}{\hbar }\Delta _{e}\mathrm{Im}[\sigma _{31}(1,1)-\allowbreak \sigma
_{42}(1,1)]\sim e\Gamma  \label{eq:second}
\end{equation}%
This contribution is of the same order of magnitude as that from the states
of subspaces (II).

In opposite case, $\hbar \left\vert J\right\vert >>\Delta _{e}$, $\cos 2\tau
\approx 1$ (again we use $J<0$ -as in J-aggregates) and $\sin 2\tau \approx
0 $. In this case only the first term in (\ref{eq:current_final1})
contributes to the current from the states of subspace (III). For this case
we get $E_{2}\approx E_{4}\approx \varepsilon _{e}$ and $E_{1,3}\approx
\varepsilon _{e}\mp J\hbar $. This leads to
\begin{equation}
\frac{2e}{\hbar }\Delta _{e}\mathrm{Im}[\sigma _{32}(1,1)+\sigma
_{41}(1,1)]\sim e\Gamma \frac{\Delta _{e}}{J\hbar }  \label{eq:first}
\end{equation}%
This contribution is much smaller than that of Eqs.(\ref{eq:third}) and (\ref%
{eq:second}), since the hopping matrix element $\Delta _{e}$ is much smaller
than the splitting between states $3$ and $2$, and states $4$ and $1$ due to
the exciton interaction (indices $1,2,3$ and $4$ label the the eigenstates
of the wire Hamiltonian in subspace III). This can cause the value of the
total current to decrease. In other words, the transitions $3\rightarrow 2$
and $4\rightarrow 1$ do not participate in electron transfer due to their
large splitting for $\hbar \left\vert J\right\vert >>\Delta _{e}$. This is
in a sense \textquotedblright exciton blocking\textquotedblright\ of
electron transmission through the bridge.

\begin{figure}[tbp]
\begin{center}
\includegraphics[width=7.cm,clip,angle=0]
{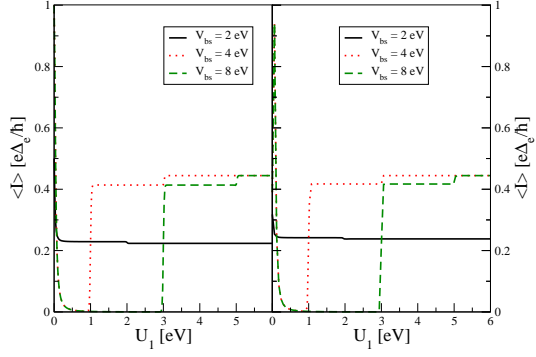}
\end{center}
\caption{The current $\langle I\rangle $ as a function of the Coulomb
interaction parameter $U_{1}=U_{2}$. $V_{bs}=2.0eV$ (solid line); $%
V_{bs}=4.0eV$ (dashed line); $V_{bs}=8eV$ (dotted+dashed line); $J=0.0eV$
(left panel) and $J=-0.05eV$ (right panel). }
\label{fig:b2040uc}
\end{figure}

Next we turn to situations where electron-electron interaction is taken into
account. Fig.\ref{fig:b2040uc} shows the current $\langle I\rangle $ as a
function of the Coulomb interaction parameter $U_{1}=U_{2}$. Fig.\ref%
{fig:biasj} depicts the current $\langle I\rangle $ as a function of the
bias voltage $V_{bs}$ for different values of the exciton coupling $J$ for
the case of non-interacting electrons as well as for the case of infinite
on-site interaction between electrons. The \textquotedblleft exciton
blocking\textquotedblright\ effect seen for non-interacting electrons
(smaller current for larger $\left\vert J\right\vert $) disappears in the
case of Coulomb blocking.

\begin{figure}[tbp]
\begin{center}
\includegraphics[width=7.cm,clip,angle=0]
{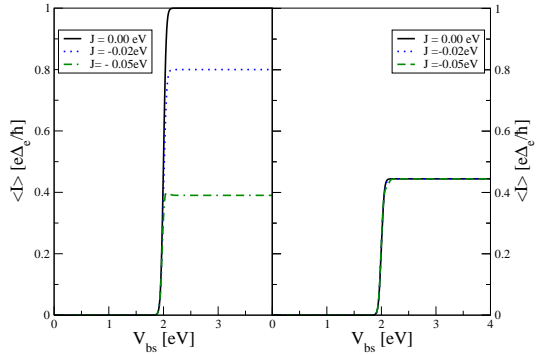}
\end{center}
\caption{The current $\langle I\rangle $ as a function of the bias voltage $%
V_{bs}$ shown for different values of the exciton coupling parameter: $J=0.0$
(solid line), $J=-0.02eV$ (dotted line), $J=-0.05eV$ (dash-dotted line).
Left panel - noninteracting electrons. Right panel $U_{1}$,$U_{2}=\infty $.}
\label{fig:biasj}
\end{figure}
This is supported by Eq.(\ref{eq:currentCB}) that does not show a direct
contribution of the states of subspace (III) to the current, and the above
evaluation of the term $\frac{2e}{\hbar }\Delta _{e}\mathrm{Im}\sigma
_{-+}(1,0)$. The point is that in the case of interacting electrons,
subspace (III) includes only states, which are acted upon exciton
interaction (see Fig.\ref{fig:subspaces}). Moreover, in the case of Coulomb
blocking, the effect of exciton-induced current exists (Sec.\ref%
{subsec:exciton_induced_current}).

\begin{figure}[tbp]
\begin{center}
\includegraphics[width=7.cm,clip,angle=0]
{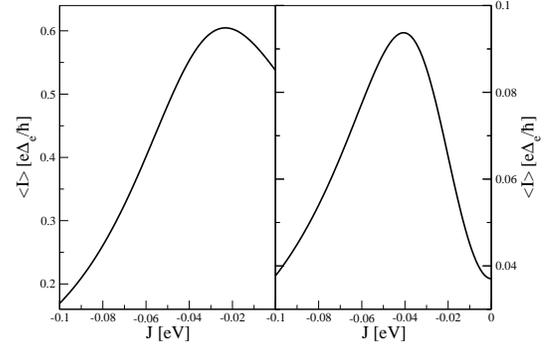}
\end{center}
\caption{The current $\langle I\rangle $ plotted against the exciton
coupling parameter for bias $V_{bs}=4.0eV$ for different energies in the $e$%
-channel: $\protect\varepsilon _{1e}=1.95eV$ and $\protect\varepsilon %
_{2e}=2.05eV$. $\Delta _{g}=0.01eV$ (left panel), $\Delta _{g}=0$ (right
panel).}
\label{fig:b40jplusgap}
\end{figure}

\begin{figure}[tbp]
\begin{center}
\includegraphics[width=7.cm,clip,angle=0]
{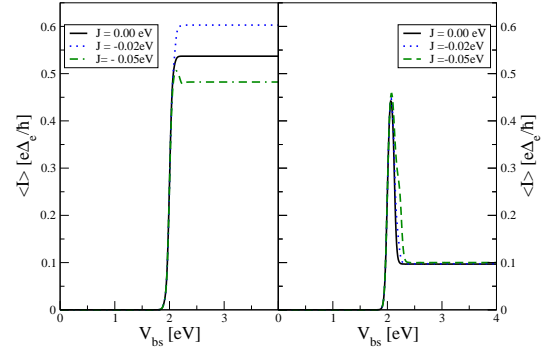}
\end{center}
\caption{Same as Fig.~\protect\ref{fig:biasj} except that $\protect%
\varepsilon _{1e}=1.95eV$ and $\protect\varepsilon _{2e}=2.05eV$.}
\label{fig:biasjgap}
\end{figure}

The effect of \textquotedblleft exciton blocking\textquotedblright\ depends
on the energy detuning $\varepsilon _{2f}-\varepsilon _{1f}$ in channel $"f"$
for a heterodimer bridge. Figs.\ref{fig:b40jplusgap} and \ref{fig:biasjgap}
show the current $\langle I\rangle $ as a function of $J$ for $\varepsilon
_{2e}-\varepsilon _{1e}=0.1eV$. $\langle I\rangle $ is seen to increase for
small $\left\vert J\right\vert $, then to decrease as $\left\vert
J\right\vert $ becomes larger. This can be related to the modification of
resonance conditions when $\varepsilon _{2e}-\varepsilon _{1e}\neq 0$.

\begin{figure}[tbp]
\begin{center}
\includegraphics[width=7.cm,clip,angle=0]
{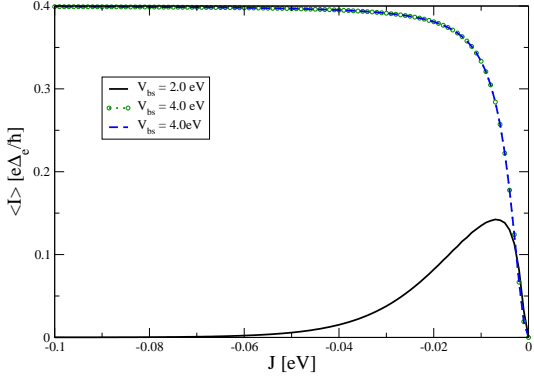}
\end{center}
\caption{Current as a function of $\left\vert J\right\vert $ for the initial
population of state $|0_{1g},0_{2g},1_{1e},0_{2e}\rangle $ equal to $1$. $%
\Delta _{g}=0.0$, $\Delta _{e}=0.01eV$, $\Gamma _{1g}=\Gamma _{2g}=0.02eV$, $%
\Gamma _{1e}=\Gamma _{2e}=0$. $V_{bs}=2.0eV$ (solid line), $V_{bs}=4.0eV$
(circles - numerical simulations, dashed - calculations with Eq.(\protect\ref%
{eq:nonLandauer_current_noRWA})).}
\label{fig:CoulonBlocking}
\end{figure}

\begin{figure}[tbp]
\begin{center}
\includegraphics[width=7.cm,clip,angle=0]
{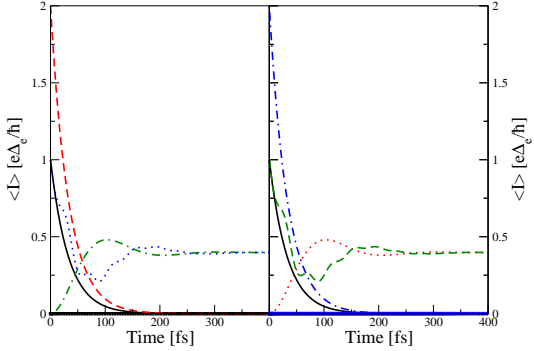}
\end{center}
\caption{Current $\langle I\rangle $ as a function of time for different
initially populated many-electron states: $|0_{1g},0_{2g},0_{1e},0_{2e}%
\rangle $ - solid, $|1_{1g},0_{2g},0_{1e},0_{2e}\rangle $ - dashed, $%
J=-0.05eV$ and $V_{bs}=8.0eV$. Left panel: $\Delta _{g}=0$, $\Delta
_{e}=0.01eV$, $\Gamma _{1g}=\Gamma _{2g}=0.02eV$, $\Gamma _{1e}=\Gamma
_{2e}=0$, $|0_{1g},0_{2g},0_{1e},1_{2e}\rangle $ - dot-dashed, $%
|0_{1g},1_{2g},0_{1e},0_{2e}\rangle $ - squares, $%
|0_{1g},0_{2g},1_{1e},0_{2e}\rangle $ - dotted. Right panel: $\Delta
_{g}=0.01eV$, $\Delta _{e}=0$, $\Gamma _{1g}=\Gamma _{2g}=0$, $\Gamma
_{1e}=\Gamma _{2e}=0.02eV,$ $|0_{1g},0_{2g},1_{1e},0_{2e}\rangle $ -
dot-dashed, $|0_{1g},1_{2g},0_{1e},0_{2e}\rangle $ - dotted, $%
|0_{1g},0_{2g},0_{1e},1_{2e}\rangle $ - squares.}
\label{fig:rho}
\end{figure}

\begin{figure}[tbp]
\begin{center}
\includegraphics[width=7.cm,clip,angle=0]
{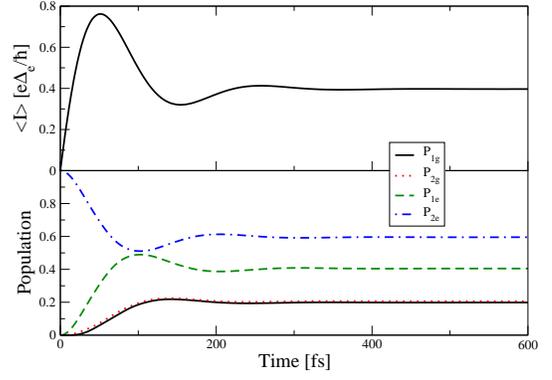}
\end{center}
\caption{Current and one-particle populations $P_{nf}=Tr(\hat{c}_{nf}^{+}%
\hat{c}_{nf^{\prime }}\protect\sigma )$ as functions of time for the initial
population of state $|0_{1g},0_{2g},0_{1e},1_{2e}\rangle $ equal to $1$. $%
J=-0.05eV$, $V_{bs}=8.0eV$, $\Delta _{g}=0.0$, $\Delta _{e}=0.01eV$, $\Gamma
_{1g}=\Gamma _{2g}=0.01eV$, $\Gamma _{M,1e}=\Gamma _{2e}=0.0$.}
\label{fig:4k3}
\end{figure}

\begin{figure}[tbp]
\begin{center}
\includegraphics[width=7.cm,clip,angle=0]
{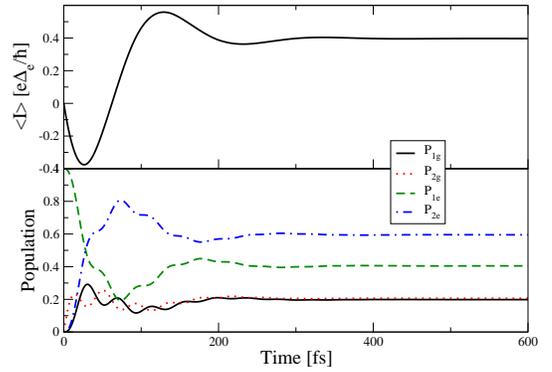}
\end{center}
\caption{Current and one-particle populations $P_{nf}=Tr(\hat{c}_{nf}^{+}%
\hat{c}_{nf^{\prime }}\protect\sigma )$ as functions of time for the initial
population of state $|0_{1g},0_{2g},1_{1e},0_{2e}\rangle $ equal to $1$. $%
J=-0.05eV$, $V_{bs}=8.0eV$, $\Delta _{g}=0.0$, $\Delta _{e}=0.01eV$, $\Gamma
_{1g}=\Gamma _{2g}=0$, $\Gamma _{1e}=\Gamma _{2e}=0.02eV$. }
\label{fig:4k4}
\end{figure}

Finally, figures \ref{fig:CoulonBlocking}, \ref{fig:rho}, \ref{fig:4k3} and %
\ref{fig:4k4} show more of the system behavior for the model with $U_{1}$,$%
U_{2}=\infty $. Fig.\ref{fig:CoulonBlocking} shows the current as a function
of $\left\vert J\right\vert $ for different values of the imposed voltage
bias $V_{bs}$. If $V_{bs}$\ is large compared to the energy difference
between the excited and ground site energies, the current behaves in
accordance with Eq.(\ref{eq:nonLandauer_current_noRWA}). If $V_{bs}$\ is
close to this energy difference, the current increases initially with $%
\left\vert J\right\vert $, and then decreases to zero. Furthermore, in
accordance with Eq.(\ref{eq:nonLandauer_current_noRWA}), the left panel of
Fig.\ref{fig:rho} shows that the steady-state current is zero for the
initial condition $\sigma _{--}(0,1)=1.$ The steady-state current is zero
also for the initial condition $\sigma (0,0)=1$, since the latter state
relaxes to $\sigma _{--}(0,1)=1$. Figures \ \ref{fig:4k3} and \ref{fig:4k4}
show the time dependence of the current and one-particle populations for
different initial conditions corresponding to the absence of relaxation in $%
e $-channel and $g$-channel, respectively.

\section{Conclusion}

\label{sec:Conclusion}

We have developed a theory of electron transport through a molecular wire in
the presence of the effect of dipolar energy-transfer interaction between
the sites in the wire. We found that such interaction, which leads to
exciton excitations in the wire, cannot in general be disregarded. We used a
model comprising a two two-level sites bridge connecting free electron
reservoirs. Expanding the density operator in the many-electron eigenstates
of the uncoupled sites, we obtain a $16\times 16$ density matrix in the
bridge subspace whose dynamics is governed by Liuoville equation that takes
into account interactions on the bridge as well as electron injection and
damping to and from the leads. Our consideration has been considerably
simplified by using the pseudospin description based on the symmetry
properties of Lie group SU(2). We studied the influence of the bias voltage,
the Coulomb repulsion and the energy-transfer interactions on the
steady-state current and in particular focus on the effect of the excitonic
interaction between bridge sites. Our calculations show that in the case of
non-interacting electrons this interaction leads to reduction in the current
at high voltage for a homodimer bridge. This effect can be called
\textquotedblleft exciton\textquotedblright blocking. The effect of
\textquotedblleft exciton\textquotedblright blocking is modified for a
heterodimer bridge, and disappears for strong Coulomb repulsion at sites. In
the latter case the exciton type interactions can open new channels for
electronic conduction. In particular, in the case of strong Coulomb
repulsion, conduction exists even when the electronic connectivity as
defined above does not exist.

To end this discussion we note that in this work we have investigated a
molecular bridge connecting metal leads. It is worthy to note that the
geometry considered could modify the effect of dipolar energy-transfer
interaction between the sites in the wire \cite{Gersten_NitzanCPL84}. This
issue will be considered elsewhere.

\textbf{Acknowledgement}

This work was supported by the German-Israeli Fund (PH, SK and AN), European
Research Comission and the Israel Science Foundation (AN), the Israel-US
binational Science Foundation (AN and BF), the Russia-Israel Scientific
Research Cooperation (BF), the Deutsche Forschungsgemeinschaft through SPP
1243 and the German Excellence Initiative via the \textquotedblleft
Nanosystems Initiative Munich (NIM)\textquotedblright\ (PH, SK and BF).

\section{Appendix A. Non-interacting electrons at a site}

\label{sec:non-interacting electrons}

The unitary transformations $\hat{Y}(\lambda _{e},\lambda _{g})=I$ \ for
subspaces (I). As to subspaces (II), Hamiltonian corresponding to the second
line of the RHS of Eq.(\ref{eq:H_wire_upBloch_lambda_f1}) where $\lambda
_{f}=1$ $\neq \lambda _{f^{\prime }}$ ($f^{\prime }\neq f$) can be
diagonalized, using the unitary transformation

\begin{equation}
\left(
\begin{array}{c}
R_{1}^{f} \\
R_{2}^{f} \\
R_{3}^{f}%
\end{array}%
\right) =\hat{T}^{f}\left(
\begin{array}{c}
r_{1}^{f} \\
r_{2}^{f} \\
r_{3}^{f}%
\end{array}%
\right) \equiv \left(
\begin{array}{ccc}
\cos 2\vartheta _{f} & 0 & -\sin 2\vartheta _{f} \\
0 & 1 & 0 \\
\sin 2\vartheta _{f} & 0 & \cos 2\vartheta _{f}%
\end{array}%
\right) \left(
\begin{array}{c}
r_{1}^{f} \\
r_{2}^{f} \\
r_{3}^{f}%
\end{array}%
\right)  \label{eq:Bloch_vector3}
\end{equation}%
where
\begin{align}
\cos 2\vartheta _{f}& =\frac{\varepsilon _{2f}-\varepsilon _{1f}}{\sqrt{%
(\varepsilon _{2f}-\varepsilon _{1f})^{2}+4\Delta _{f}^{2}}},\text{ }  \notag
\\
\sin 2\vartheta _{f}& =\frac{-2\Delta _{f}}{\sqrt{(\varepsilon
_{2f}-\varepsilon _{1f})^{2}+4\Delta _{f}^{2}}}  \label{eq:sin_cos_teta}
\end{align}%
The matrix elements of $\hat{T}^{f}$ are connected with the unitary
transformations $\hat{Y}(\lambda _{e},\lambda _{g})$ for subspaces (II) by
formula $T_{nj}^{f}=(1/2)Tr(\hat{\sigma}_{n}\hat{Y}^{+}\hat{\sigma}_{j}\hat{Y%
})$ where $\hat{\sigma}_{n}$ and $\hat{\sigma}_{j}$ are Pauli matrices.

\subsection{Unitary transformations for subspaces (II)}

Consider subspaces (II). In the limit $U_{m}=0$, the matrix $\hat{T}$ $^{f}$%
, Eq.(\ref{eq:Bloch_vector3}), with matrix elements $T_{nj}^{f}=(1/2)Tr[\hat{%
\sigma}_{n}\hat{Y}^{+}(\lambda _{f}=1;\lambda _{f^{\prime }}=0,2)\hat{\sigma}%
_{j}\hat{Y}(\lambda _{f}=1;\lambda _{f^{\prime }}=0,2)]$ describes a
rotation by mixing angle $2\vartheta _{f}$ around axis $\ "y"$. $\hat{Y}%
(\lambda _{f}=1;\lambda _{f^{\prime }}=0,2)$ is an unitary operator defined
by
\begin{equation}
\hat{Y}^{+}(\lambda _{f}=1;\lambda _{f^{\prime }}=0,2)=\left(
\begin{array}{cc}
\cos \vartheta _{f} & \sin \vartheta _{f} \\
-\sin \vartheta _{f} & \cos \vartheta _{f}%
\end{array}%
\right)  \label{eq:Psi_f+-single}
\end{equation}%
which enables us to obtain eigenstates%
\begin{align}
\left(
\begin{array}{c}
\Phi _{+}(\lambda _{f}=1;\lambda _{f^{\prime }}=0,2) \\
\Phi _{-}(\lambda _{f}=1;\lambda _{f^{\prime }}=0,2)%
\end{array}%
\right) & =\hat{Y}^{+}(\lambda _{f}=1;\lambda _{f^{\prime }}=0,2)  \notag \\
\times \hat{\chi}(\lambda _{f}& =1;\lambda _{f^{\prime }}=0,2)
\label{eq:Psi_f+-1}
\end{align}%
and eigenvalues%
\begin{align}
E_{\pm }(\lambda _{f}& =1;\lambda _{f^{\prime }}=0,2)=\frac{1}{2}[\lambda
_{e}(\varepsilon _{1e}+\varepsilon _{2e})+(\varepsilon _{2f}-\varepsilon
_{1f})  \notag \\
& \pm \sqrt{(\varepsilon _{2f}-\varepsilon _{1f})^{2}+4\Delta _{f}^{2}}]
\label{eq:eigenvalues_noncollective1}
\end{align}%
for subspaces (II). Here the many-electron eigenstates of the uncoupled
sites are given by $\hat{\chi}(1,0)=\left(
\begin{array}{c}
|0_{1g},0_{2g},0_{1e},1_{2e}\rangle \\
|0_{1g},0_{2g},1_{1e},0_{2e}\rangle%
\end{array}%
\right) $, $\hat{\chi}(0,1)=\left(
\begin{array}{c}
|0_{1g},1_{2g},0_{1e},0_{2e}\rangle \\
|1_{1g},0_{2g},0_{1e},0_{2e}\rangle%
\end{array}%
\right) $, $\hat{\chi}(1,2)=\left(
\begin{array}{c}
|1_{1g},1_{2g},0_{1e},1_{2e}\rangle \\
|1_{1g},1_{2g},1_{1e},0_{2e}\rangle%
\end{array}%
\right) $ and $\hat{\chi}(2,1)=\left(
\begin{array}{c}
|0_{1g},1_{2g},1_{1e},1_{2e}\rangle \\
|1_{1g},0_{2g},1_{1e},1_{2e}\rangle%
\end{array}%
\right) $.

Taking the expectation value of the current, Eq.(\ref{eq:current_lambda=1}),
we get%
\begin{align}
\langle I\rangle & =\frac{2e}{\hbar }\{\sum_{\lambda _{f^{\prime
}}=0,2;f}\Delta _{f}\mathrm{Im}\sigma _{-+}(\lambda _{f}=1;\lambda
_{f^{\prime }})-\sum_{\alpha \beta }\mathrm{Im}\sigma _{\beta \alpha }(1,1)
\notag \\
& \times \lbrack \hat{Y}^{+}(1,1)\tilde{\chi}^{+}(1,1)(\Delta
_{e}b_{e}+\Delta _{g}b_{g})\tilde{\chi}(1,1)\hat{Y}(1,1)]_{\alpha \beta }\}
\label{eq:current}
\end{align}%
where we put $r_{2}^{f}=R_{2}^{f}$ for $\lambda _{f}=1$ and $\lambda
_{f^{\prime }}=0$,$2$ that follows from Eq.(\ref{eq:Bloch_vector3}) and used
$\langle R_{2}^{f}(\lambda _{f}=1;\lambda _{f^{\prime }}=0,2)\rangle =Tr(%
\hat{\sigma}_{2}\sigma )=2\mathrm{Im}\sigma _{-+}(\lambda _{f}=1;\lambda
_{f^{\prime }}=0,2)$. Indices "$+$" and "$-$" in Eq.(\ref{eq:current})
correspond to the functions $\Phi _{+}(1,\lambda _{g})$ and $\Phi
_{-}(1,\lambda _{g})$, respectively, in Table I.

\subsection{Unitary transformation for subspace (III)}

The calculation of $\hat{Y}^{+}(1,1)$ is more involved. Consider for brevity
a homodimer bridge with $\varepsilon_{ng}=0$, $\varepsilon_{ne}=%
\varepsilon_{e}$ and $\Delta_{g}=0$. Bearing in mind future generalizations
of our model to $N$-sites, we shall transform the Paulion operators ($%
b_{f}^{+},b_{f}$) to fermion operators ($\beta_{f}^{+},\beta_{f}$) through
the Jordan-Wigner transformation \cite{Fra98,Che63,Spano91}:

\begin{align}
& \beta _{e}=b_{e},\text{ }\beta _{e}^{+}=b_{e}^{+},\beta _{g}=\exp \left(
i\pi b_{e}^{+}b_{e}\right) b_{g},  \notag \\
& \text{ }\beta _{g}^{+}=b_{g}^{+}\exp \left( -i\pi b_{e}^{+}b_{e}\right)
\label{eq:JW}
\end{align}%
Then $\hat{H}_{wire}$, Eqs.(\ref{eq:H_wire}) and (\ref{eq:H_wire_up}), can
be rewritten for subspace (III) in terms of the fermion operators as
\begin{equation}
\hat{H}_{wire}(\lambda _{e}=\lambda _{g}=1)=\varepsilon _{e}-\Delta
_{e}(\beta _{e}^{+}+\beta _{e})-\hbar J(\beta _{e}^{+}\beta _{g}+\beta
_{g}^{+}\beta _{e})  \label{eq:H_wire_Fermi_lambda=1}
\end{equation}%
Eq.(\ref{eq:H_wire_Fermi_lambda=1}) \ is a quadratic in Fermi operators and
can be diagonalized in two stages. Its \ \textquotedblleft excitonic" part $%
\hat{H}_{ex}=-\hbar J(\beta _{e}^{+}\beta _{g}+\beta _{g}^{+}\beta _{e})$ is
readily transformed to satisfy the condition $\hat{H}_{ex}=\sum_{j}\hbar
\epsilon _{j}a_{j}^{+}a_{j}$ if we take \cite{Che63}%
\begin{align}
a_{j}& =\sqrt{2/3}(\beta _{g}\sin \frac{\pi j}{3}+\beta _{e}\sin \frac{2\pi j%
}{3}),  \notag \\
\epsilon _{j}& =-2J\cos \frac{\pi j}{3},\text{\ \ \ }j=1,2
\label{eq:Fermion}
\end{align}%
where $a_{j}$ are also Fermi operators. The corresponding occupation number
basis set contains $2^{2}=4$ eigenfunctions of the system. The
single-excited states are given by $a_{j}^{+}|0\rangle =\sum\limits_{f}\phi
_{jf}|f\rangle =\sqrt{1/2}(|g\rangle +(-1)^{j-1}|e\rangle )$ where $%
|0\rangle \equiv |1_{1g},0_{2g},1_{1e},0_{2e}\rangle $ is the \ "vacuum"
state, and $|g\rangle \equiv |0_{1g},1_{2g},1_{1e},0_{2e}\rangle $ and $%
|e\rangle \equiv |1_{1g},0_{2g},0_{1e},1_{2e}\rangle $ are the states with
the corresponding donor acceptor pair $\ $excited. The eigenstate with two
excitations can be written down in terms of the Slater determinant

\begin{equation}
a_{j_{1}}^{+}a_{j_{2}}^{+}|0\rangle =\left\vert
\begin{array}{cc}
\phi _{j_{1}e} & \phi _{j_{1}g} \\
\phi _{j_{2}e} & \phi _{j_{2}g}%
\end{array}%
\right\vert |eg\rangle =\frac{1}{2}[(-1)^{j_{2}}-(-1)^{j_{1}}]|eg\rangle
\label{eq:two-particle state_eg}
\end{equation}%
with energy $\epsilon _{1}+\epsilon _{2}=0$ equal to that of the vacuum
state where $|eg\rangle \equiv |0_{1g},1_{2g},0_{1e},1_{2e}\rangle $. The
wire Hamiltonian can be written down in terms of $a_{j}$ as $\hat{H}%
_{wire}(\lambda _{e}=\lambda _{g}=1)=\varepsilon _{e}+\sum_{j}\hat{H}_{j}$
where $\hat{H}_{j}=\hat{F}_{j}+(-1)^{j}\hbar Ja_{j}^{+}a_{j}$, $\hat{F}%
_{j}=(-1)^{j}(\Delta _{e}/\sqrt{2})(a_{j}^{+}+a_{j})$ is the \ "hopping"
operator with the only nonzero matrix elements involving states which differ
by a single excitation: $\langle 0|\hat{F}_{j}a_{j}^{+}|0\rangle
=(-1)^{j}\Delta _{e}/\sqrt{2},$ $\langle 0|a_{j_{2}}\hat{F}%
_{j_{1}}a_{j_{1}}^{+}a_{j_{2}}^{+}|0\rangle =(-1)^{j_{1}}\Delta _{e}/\sqrt{2}
$. The eigenstates and eigenvalues of $\hat{H}_{wire}(\lambda _{e}=\lambda
_{g}=1)$ can be calculated now as follows. $\hat{\Phi}(1,1)=Y^{+}(1,1)\hat{%
\chi}(1,1)$ where $Y^{+}(1,1)$ is given by Eqs.(\ref{eq:Y(1,1)}) and (\ref%
{eq:cos_sin}), $\hat{\chi}(1,1)=\left(
\begin{array}{c}
|0\rangle \\
|e\rangle \\
|g\rangle \\
|eg\rangle%
\end{array}%
\right) ,$ and

\begin{align}
\hat{\Phi}(1,1)&=\frac{1}{\sqrt{2}}\left(
\begin{array}{c}
(|0\rangle+|eg\rangle)\sin\tau-(|e\rangle+|g\rangle)\cos\tau \\
(|e\rangle+|g\rangle)\sin\tau+(|0\rangle+|eg\rangle)\cos\tau \\
(|e\rangle-|g\rangle)\cos\tau+(|0\rangle-|eg\rangle)\sin\tau \\
(|0\rangle-|eg\rangle)\cos\tau-(|e\rangle-|g\rangle)\sin\tau%
\end{array}
\right)  \notag \\
&\equiv\left(
\begin{array}{c}
|\Phi_{1}\rangle \\
|\Phi_{2}\rangle \\
|\Phi_{3}\rangle \\
|\Phi_{4}\rangle%
\end{array}
\right)  \label{eq:Fi}
\end{align}
%

Substituting Eq.(\ref{eq:Y(1,1)}) into Eq.(\ref{eq:current}) for the
current, we get Eq.(\ref{eq:current_final1}) for $\Delta _{g}=0$.

\section{Appendix B. Unitary transformation for subspace (III) for
interacting electrons at a site}

In the limit of strong Coulomb repulsion, the operator $\hat{Y}^{+}(1,1)$ is
reduced to that defined by Eq.(\ref{eq:Y+(1,1)blockade}) in accordance with
Eqs.(\ref{eq:Fermion}), since the\ "hopping" operator $\hat{F}%
_{j}=(-1)^{j}(\Delta _{e}/\sqrt{2})(a_{j}^{+}+a_{j})$ has no nonzero matrix
elements involving states with a single excitation $|e\rangle $ and $%
|g\rangle $ (see Appendix A).

Substituting Eq.(\ref{eq:Y+(1,1)blockade}) into Eq.(\ref{eq:current}) for
the current, we get Eq.(\ref{eq:currentCB}) for $\Delta _{g}=0$.

\section{Appendix C}

The steady-state solution of Eqs. (\ref{eq:density_matrix}), (\ref%
{eq:sigma_albe}) in the RWA approximation gives for the case under
consideration

\begin{equation}
\sigma (0,0)=\sigma (0,2)=\sigma _{++}(0,1)=\sigma _{+-}(0,1)=\sigma
_{-+}(0,1)=0  \label{eq:sigma=0}
\end{equation}%
and $\sigma _{--}(0,1)$ and $\sigma (2,0)$ are arbitrary. Putting $\sigma
_{--}(0,1)=\sigma (2,0)=0$, we get
\begin{equation}
\sigma _{-+}(1,0)=\frac{i\hbar }{-8\Delta _{e}}\{\Gamma _{M,1g}Tr\sigma
(1,0)+\Gamma _{M,2g}Tr\sigma (1,1)\}  \label{eq:sigma_-+}
\end{equation}%
and $Tr\sigma (1,0)=(\Gamma _{M,2g}/\Gamma _{M,1g})Tr\sigma (1,1)$. Then
using the normalization condition
\begin{equation}
Tr\sigma (1,0)+Tr\sigma (1,1)=1  \label{eq:normCBandG_e=0}
\end{equation}%
and Eq.(\ref{eq:currentCB}), we obtain Eq.(\ref{eq:nonLandauer_current}) of
Sec.\ref{subsec:exciton_induced_current}.


\begin{thebibliography}{10}

\bibitem{nitz03a}
A. Nitzan and M.~A. Ratner, Science {\bf 300},  1384  (2003).

\bibitem{Nitzan08Science}
M. Galperin, M.~A. Ratner, A. Nitzan, and A. Troisi, Science {\bf 319},  1056
  (2008).

\bibitem{Chen09}
F. Chen and N.~J. Tao, Accounts of Chemical Research {\bf 42},  429  (2009).

\bibitem{Heath09}
J.~R. Heath, Annual Review of Materials Research {\bf 39},  1  (2009).

\bibitem{Lang00}
M.~D. Ventra, S.~T. Pantelides, and N.~D. Lang, Phys. Rev. Lett. {\bf 84},  979
   (2000).

\bibitem{Schon02}
J. Heurich, J.~C. Cuevas, W. Wenzel, and G. Schon, Phys. Rev. Lett. {\bf 88},
  256803  (2002).

\bibitem{Evers04}
F. Evers, F. Weigend, and M. Koentopp, Phys. Rev. B {\bf 69},  235411  (2004).

\bibitem{Ratner02CP}
Y. Xue, S. Datta, and M.~A. Ratner, Chem. Phys. {\bf 281},  151  (2002).

\bibitem{Ghosh02CP}
P. Damle, A.~W. Ghosh, and S. Datta, Chem. Phys. {\bf 281},  171  (2002).

\bibitem{Kohler05}
S. Kohler, J. Lehmann, and P. Hanggi, Phys. Reports {\bf 406},  379  (2005).

\bibitem{Kaiser06}
F.~J. Kaiser, M. Strass, S. Kohler, and P. Hanggi, Chem. Phys. {\bf 322},  193
  (2006).

\bibitem{Meir92}
Y. Meir and N.~S. Wingreen, Phys. Rev. Lett. {\bf 68},  2512  (1992).

\bibitem{Nitzan07JPhys}
M. Galperin, M.~A. Ratner, and A. Nitzan, J. Phys.: Condens. Matter {\bf 19},
  103201  (2007).

\bibitem{Dav71}
A.~S. Davydov, {\em Theory of Molecular Excitons} (Plenum, New York, 1971).

\bibitem{Agranovich68}
V.~M. Agranovich, {\em Teorija Excitonov} (Nauka, Moscow, 1968).

\bibitem{Craig68}
D.~P. Craig and S.~H. Walmsley, {\em Excitons in molecular crystals} (Benjamin,
  New York, 1968).

\bibitem{Agranovich77}
V.~M. Agranovich and A.~A. Zakhidov, Chem. Phys. Lett. {\bf 50},  278  (1977).

\bibitem{Agranovich00}
M. Hoffmann {\it et~al.}, Chem. Phys. {\bf 258},  73  (2000).

\bibitem{Reineker09}
C. Warns, I. J.Lalov, and P. Reineker, Journal of Luminescence {\bf 129},  1840
   (2009).

\bibitem{Thorne90}
J.~R.~G. Thorne, S.~T. Repinec, S.~A. Abrash, and R.~M. Hochstrasser, Chem.
  Phys. {\bf 146},  315  (1990).

\bibitem{Trommsdorff92}
A. Tilgner, H.~P. Trommsdorff, J.~M. Zeigler, and R.~M. Hochstrasser, J.
  Chem.Phys. {\bf 96},  781  (1992).

\bibitem{Shimizu00}
M. Shimizu {\it et~al.}, J. Luminescence {\bf 87--89},  933  (2000).

\bibitem{BennistonJACS05}
A.~C. Benniston, A. Harriman, P. Li, and C.~A. Sams, J. Am. Chem. Soc. {\bf
  127},  2553  (2005).

\bibitem{KimJACS06}
O.-K. Kim, J. Je, and J.~S. Melinger, J. Am. Chem. Soc. {\bf 128},  4532
  (2006).

\bibitem{Tang06}
J. Tang, Y. Wang, C. Nuckolls, and S.~J. Wind, J. Vac. Sci. Technol. B {\bf
  24},  3227  (2006).

\bibitem{Gersten_NitzanCPL84}
J.~I. Gersten and A. Nitzan, Chem. Phys. Lett. {\bf 104},  31  (1984).

\bibitem{Gersten_NitzanJCP85}
X.~M. Hua, J.~I. Gersten, and A. Nitzan, J. Chem. Phys. {\bf 83},  3650
  (1985).

\bibitem{Polman07}
H. Mertens, A.~F. Koenderink, and A. Polman, Phys. Rev. B {\bf 76},  115123
  (2007).

\bibitem{Polman09}
H. Mertens and A. Polman, J. Appl. Phys. {\bf 105},  044302  (2009).

\bibitem{Wiederrecht04}
G.~P. Wiederrecht, G.~A. Wurtz, and J. Hranisavljevic, Nanoletters {\bf 4},
  2121  (2004).

\bibitem{Wurtz07}
G.~A. Wurtz {\it et~al.}, Nanoletters {\bf 7},  1297  (2007).

\bibitem{Cade09}
N.~I. Cade, T. Ritman-Meer, and D. Richards, Phys. Rev. B {\bf 79},  241404(R)
  (2009).

\bibitem{Bondarev09}
I. Bondarev, L.~M. Woods, and K. Tatur, Phys. Rev. B {\bf 80},  085407  (2009).

\bibitem{Schatz03JPCB}
K.~L. Kelly, E. Coronado, L.~L. Zhao, and G.~C. Schatz, J. Phys. Chem. B {\bf
  107},  668  (2003).

\bibitem{Stockman03}
K. Li, M.~I. Stockman, and D.~J. Bergman, Phys. Rev. Lett. {\bf 91},  227402
  (2009).

\bibitem{Markel05}
V.~A. Markel, J. Phys. B: At. Mol. Opt. Phys. {\bf 38},  L347  (2005).

\bibitem{Wang_Shen06}
F. Wang and Y.~R. Shen, Phys. Rev. Lett. {\bf 97},  206806  (2006).

\bibitem{Brixner06}
T. Brixner {\it et~al.}, Phys. Rev. B {\bf 73},  125437  (2006).

\bibitem{Sukharev_Seideman07}
M. Sukharev and T. Seideman, J. Phys. B: At. Mol. Opt. Phys. {\bf 40},  S283
  (2007).

\bibitem{Nitzan06PRL}
M. Galperin, A. Nitzan, and M.~A. Ratner, Phys. Rev. Lett. {\bf 96},  166803
  (2006).

\bibitem{Fra98}
E. Fradkin, {\em Field theories of condensed matter systems} (Addison-Wesley,
  New York, 1991).

\bibitem{Schreiber06}
S. Welack, M. Schreiber, and U. Kleinekathofer, J. Chem. Phys. {\bf 124},
  044712  (2006).

\bibitem{Novotny02}
T. Novotny, Europhys. Lett. {\bf 59},  648  (2002).

\bibitem{All75}
L. Allen and J.-H. Eberly, {\em Optical resonance and two-level atoms} (John
  Wiley \& Sons, New York, London, Sydney, Toronto, 1975).

\bibitem{Hio81}
F.~T. Hioe and J.~H. Eberly, Phys. Rev. Lett. {\bf 47},  838  (1981).

\bibitem{yang-etal.04}
Z.~S. Yang, N.~H. Kwong, and R. Binder, Phys. Rev. B {\bf 70},  195319
  (2004).

\bibitem{Bryllert02}
T. Bryllert {\it et~al.}, Appl. Phys. Letters {\bf 80},  2681  (2002).

\bibitem{Brandes08PRB}
R. Sanchez, G. Platero, and T. Brandes, Phys. Rev. B {\bf 78},  125308  (2008).

\bibitem{Che63}
D.~B. Chesnut and A. Suna, J. Chem. Phys. {\bf 39},  146  (1963).

\bibitem{Spano91}
F.~C. Spano, Phys. Rev. Lett. {\bf 24},  3424  (1991).

\end{thebibliography}

\end{document}